\documentclass[12pt,a4paper]{article}
\usepackage{graphicx}
\usepackage{amssymb}
\usepackage{amsmath}
\usepackage{bm}
\usepackage{color}
\usepackage{theorem}
\usepackage{subfigure}

\usepackage[sort&compress,numbers, merge]{natbib}

\definecolor{Orange}{cmyk}{0,0.61,0.87,0}
\definecolor{JungleGreen}{cmyk}{0.99,0,0.52,0}
\definecolor{OliveGreen}{cmyk}{0.64,0,0.95,0.40}
\definecolor{Brown}{cmyk}{0,0.81,1,0.60}
\definecolor{RoyalBlue}{cmyk}{0.71,0.53,0,0.12}

\newcommand{\Min}{\ensuremath{M_{\rm in}}}
\newcommand{\MGUT}{\ensuremath{M_{\rm GUT}}}

\allowdisplaybreaks[1]

\usepackage[colorlinks=true, linkcolor=OliveGreen, citecolor=RoyalBlue,
urlcolor=Brown]{hyperref}

\begin{document}

\begin{titlepage}
\pagestyle{empty}
%\baselineskip=21pt
%\rightline{\tt astro-ph/yymmnnn}
{\tt
\rightline{KCL-PH-TH/2018-03, CERN-TH/2018-015, KIAS-P18010}
\rightline{IPMU18-0023, UMN-TH-3709/18, FTPI-MINN-18/01}
}
\vspace{0.1cm}
\begin{center}
{\bf {\LARGE
Stop Coannihilation in the CMSSM
\vskip +0.2cm
and SubGUT Models
} }
\end{center}

\vspace{0.05cm}
\begin{center}
%\vskip +0.4in
{\bf John~Ellis}~$^{1}$,
{\bf Jason L. Evans}~$^2$,
 {\bf Feng Luo}~$^3$\\
\vskip 0.1in
{\bf Keith~A.~Olive}~$^{4}$, {\bf Jiaming Zheng}~$^5$
\vskip 0.2in
{\small {\it
$^1${Theoretical Particle Physics and Cosmology Group, Department of Physics, \\ King's College London, Strand, London~WC2R~2LS, UK; \\
    National Institute of Chemical Physics and Biophysics, R{\" a}vala 10, 10143 Tallinn, Estonia; \\
Theoretical Physics Department, CERN, CH--1211 Geneva 23, Switzerland}\\
\vspace{0.2cm}
$^2${School of Physics, KIAS, Seoul 130-722, Korea}\\
\vspace{0.2cm}
$^3${Kavli IPMU (WPI), UTIAS, The University of Tokyo, Kashiwa, Chiba 277-8583, Japan}\\
\vspace{0.2cm}
$^4${William I.\ Fine Theoretical Physics Institute, School of Physics and
 Astronomy, University of Minnesota, Minneapolis, Minnesota 55455, USA}\\
 \vspace{0.2cm}
$^5${\it Department of Physics, University of Tokyo, Bunkyo-ku, Tokyo 113--0033, Japan}\\
}}

\vspace{0.5cm}
%\clearpage
{\bf Abstract}\\

\end{center}
%\baselineskip=18pt \noindent
%%%%%%%%%%%%%%%%%%%%%%%%%%%%%%%%%%%%%%%%%%%%%%%%%
{
Stop coannihilation may bring the relic density of heavy supersymmetric dark matter particles into the range
allowed by cosmology. The efficiency of this process is enhanced by
stop-antistop annihilations into the longitudinal (Goldstone) modes of the $W$ and $Z$ bosons,
as well as by Sommerfeld enhancement of stop annihilations and the effects of bound states.
Since the couplings of the stops to the Goldstone modes are proportional to the trilinear soft supersymmetry-breaking
$A$-terms, these annihilations are enhanced when the $A$-terms are large. However,
the Higgs mass may be reduced below the measured value if the $A$-terms are too large.
Unfortunately, the interpretation of
this constraint on the stop coannihilation strip is clouded by differences between the available Higgs mass calculators.
For our study, we use as our default calculator {\tt FeynHiggs~2.13.0}, the
most recent publicly available version of this code. Exploring the CMSSM parameter space, we find that along the stop coannihilation
strip the masses of the stops are severely split by the large $A$-terms. This suppresses
the Higgs mass drastically for $\mu$ and $A_0 > 0$, whilst the extent of the stop 
coannihilation strip is limited for $A_0 < 0$ and either sign of $\mu$.  However,
in sub-GUT models, reduced renormalization-group running mitigates the effect of the
large $A$-terms, allowing larger LSP masses to be consistent with the Higgs mass calculation.
We give examples where the dark matter particle mass may reach $\gtrsim 8$~TeV. \\}

%%%%%%%%%%%%%%%%%%%%%%%%%%%%%%%%%%%%%%%%%%%%%%%%

\vfill
\leftline{January 2018}
\end{titlepage}

%\textcolor{RoyalBlue}

\section{Introduction}

Searches for supersymmetry \cite{nosusy} at the LHC have explored much of the theory space
favoured previously in the context of simplified phenomenological models with universal
soft super- symmetry-breaking parameters at an input GUT scale. However, in our opinions
supersymmetry (SUSY) remains one of the most attractive options for physics beyond the
Standard Model, since it facilitates grand unification of the gauge couplings \cite{Ellis:1990zq} , improves
the naturalness of the electroweak mass hierarchy \cite{Maiani:1979cx} and plays an essential role in string
theory. Moreover, the lightest supersymmetric particle (LSP) is an excellent cold dark matter candidate if $R$-parity is conserved \cite{ehnos},
as we assume here. In addition, supersymmetry stabilizes the electroweak vacuum \cite{Ellis:2000ig,degrassi},
can trigger electroweak symmetry breaking \cite{ewsb} and predicted successfully
the mass of a Higgs boson with couplings similar to those in the Standard Model \cite{mh,ENOS}.

Therefore we are motivated to pursue the search for supersymmetry, and note that there are still
regions of supersymmetric model space that the LHC has yet to explore, and may never
reach. Some of the regions that are difficult to see at the LHC can be seen through indirect detection,
and another promising avenue is the search for proton decay \cite{Takhistov:2016eqm}. If the theory at the GUT scale is minimal SU(5),
the Wilson coefficients of the dimension-5 proton-decay operators tend to be large, destabilizing the proton \cite{Goto:1998qg, mp}.
In general, unless $\tan\beta\lesssim 5$, the proton is unstable for a SUSY-breaking scale that can explain dark
matter~\footnote{Exceptions to this are models such as pure gravity meditation that have a large hierarchy
between the sfermion and gaugino masses \cite{evno}.} \cite{eelnos,eemno}. However, in this work we ignore such constraints,
assuming that the GUT-scale theory is either not minimal SU(5) or has some additional symmetry,
such as a Peccei-Quinn symmetry, which enhances the proton lifetime. In this way, we are not hostages of
some unknown high-scale dynamics.

Instead of worrying about constraints that are dependent on the UV completion of the model,
we take a phenomenological approach and focus on the regions of supersymmetric model space
that has not yet been probed by the LHC. Included in these unexplored regions are strips of parameter
space extending to larger masses where the thermal abundance of relic LSPs is brought down
into the range allowed by the cosmological cold dark matter density measurements~\cite{Planck} via some enhancement
of the conventional annihilation mechanism, such as rapid annihilation through
heavy Higgs bosons or coannihilation with some other, nearly-degenerate
supersymmetric particle(s) \cite{gs}.

Examples of possible coannihilation partners include sleptons \cite{efo,stau,celmov,delm}, electroweak inos \cite{chacoann},
squarks \cite{stopco,eds,eoz,interplay,raza} and gluinos \cite{glu,shafi,hari,evo,deSimone:2014pda,liantao,raza,ELO,eelo}.
Coannihilation of the LSP with the lighter stau slepton has been
explored extensively, and is now almost excluded by LHC searches \cite{celmov,delm}.
The cosmological cold dark matter density can be obtained via coannihilations with Higgsinos
if the LSP mass $\sim 1$~TeV \cite{osi}, and by coannihilations with Winos if the LSP mass $\sim 3$~TeV \cite{winomass1,winomass}.
Much larger LSP masses, and hence much heavier sparticle spectra, are possible if the
LSP coannihilates with strongly-interacting sparticles such as gluinos or stop squarks.
Coannihilation with gluinos is not possible in models with universal gaugino masses at the
GUT scale, though it is possible if this assumption is relaxed. On the other hand,
coannihilation with stop squarks is possible in models with universal
soft supersymmetry-breaking parameters, and is a scenario capable of raising
the sparticle spectrum into the multi-TeV range and evading LHC searches~\cite{interplay}.

The fact that stop coannihilation is such a promising scenario for reconciling a heavy
supersymmetric spectrum with the attractive possibility that the LSP provides the
cosmological relic density motivates the re-examination of this scenario, which we
undertake in this paper. We consider in particular, various effects that
tend to extend the stop coannihilation strip, including annihilations into
longitudinal (Goldstone) components of the $W$ and $Z$ bosons, large trilinear soft
supersymmetry-breaking $A$-terms, Sommerfeld enhancement \cite{Sommerfeld1931,hisano} of
stop annihilations \cite{stopsom,deSimone:2014pda,eoz}, and the possible effects of bound states \cite{bound,LL}.

We pay particular attention to the limitations on the stop coannihilation strip imposed by the LHC measurement of
the Higgs mass \cite{125}. The interpretation of this constraint is sensitive to details of Higgs mass calculations
with heavy sparticle spectra. These have been studied extensively recently, but still with
significant differences between the available Higgs mass calculators \cite{FH,susyhd,SSARD}. Within the CMSSM \cite{eelnos,funnel,cmssm,elos},
 in which the soft supersymmetry-breaking scalar and gaugino masses are assumed to be universal
 at the GUT scale, we find using the most recent publicly available version of the FeynHiggs Higgs
 mass calculator, {\tt FeynHiggs~2.13.0}, that the Higgs mass constraint
is a severe limitation on the size of the LSP mass.

In addition to the CMSSM, we also consider its `sub-GUT' generalization \cite{subGUT,elos,eelnos,mcsubgut} in which universality
of the soft supersymmetry-breaking parameters is imposed at a lower scale, as occurs in mirage unification models \cite{mirage}.
Sub-GUT models can have enhanced coannihilations of the LSP and the lighter stop into
final states including $W$ and $Z$ bosons. This enhancement occurs because the
masses of the left- and right-handed stop masses are more degenerate than in the CMSSM,
as a consequence of the reduced renormalization-group running.  With the masses less split,
for any fixed value of $m_{\tilde t_1}^2+m_{\tilde t_2}^2$, the ratio $A_t^2/(m_{\tilde t_1} m_{\tilde t_2})$,
important for the Higgs mass calculation, is decreased. Since the ratio $A_t^2/(m_{\tilde t_1}^2+m_{\tilde t_2}^2)$ is important for
determining the rate of stop-antistop annihilation into the longitudinal modes of the $W$ and $Z$,
the stop coannihilation strip can be extended in such sub-GUT models. Moreover,
in certain regions of the parameter space of sub-GUT models, the masses have the
special relationship $2m_{\tilde t_1} \simeq 2 m_{\chi} \simeq m_{H}$, where
$m_{\tilde t_1}$, $m_{\chi}$ and $m_H$ are the lightest stop mass, LSP mass and
heavy CP-even Higgs boson mass, respectively. In these regions,
the stop-antistop annihilation rate is enhanced by resonance effects because
$2m_{\tilde t_1}  \simeq m_{H}$, amplifying the ability of stop coannihilation
with the LSP to reduce the relic density, and the stop coannihilation strip is further extended.
We give examples of sub-GUT scenarios where the dark matter particle mass may reach $\sim 7$~TeV.

The layout of our paper is as follows. In Section~2 we discuss the impact of annihilations
into longitudinal components of the $W$ and $Z$ on the extent of the stop coannihilation strip.
In Section~3 we discuss the impact of bound states, showing how the longitudinal components
of the gauge bosons enhance the decays of the bound states. In Section~4 we illustrate the
importance of these effects in the CMSSM, discussing the potential impact of the Higgs mass
constraint. In Section~5  we extend the analysis to
sub-GUT models in which the stop masses are less split. In such a case, the Higgs mass is less suppressed and
the stop coannihilation strip may extend to larger values of the LSP mass. Finally, Section~6
summarizes our conclusions.

\section{The Goldstone Equivalence Theorem and Stop \\ Coannihilation}
%\vspace{.3in}

The Goldstone Equivalence \cite{GBET} theorem states that the longitudinal components of the gauge bosons of a broken symmetry
retain the interactions they would have in the absence of gauge interactions, i.e., they interact
as Goldstone bosons. If the interactions of the Goldstone bosons are large, they may enhance
the interactions of the gauge bosons.  The best-known example of this is $t\to W^+ b$ decay.  Naively,
one would have expected that the dominant contribution to this decay would be proportional to $g_2^2$,
where $g_2$ is the SU(2) electroweak gauge coupling, since this appears to be a weak process.
However, since the charged Goldstone boson in the Standard Model couples to $t$ and $b$ with a strength $y_t$,
where $y_t$ is the top Yukawa coupling that is larger than $g_2$, this decay is enhanced:
\begin{eqnarray}
\Gamma_t\simeq \frac{g_2^2}{64\pi} \frac{m_t^3}{m_W^2}= \frac{y_t^2}{32\pi} m_t \, .
\end{eqnarray}
Similar behaviours are present in all scattering processes involving the $W$ and $Z$.
This type of enhancement turns out to be relevant when considering coannihilation processes involving
the stop, as we see below.

Since the MSSM is a two-Higgs-doublet model, the Goldstone bosons are mixtures of states in the $H_{u,d}$
multiplets that give masses to the up- and down-type quarks:
\begin{eqnarray}
H_u\supset \sin\beta \left(\begin{array}{c} G^+\\ \frac{1}{\sqrt{2}} G^0\end{array}\right) \quad \quad H_d\supset -\cos\beta \left(\begin{array}{c}\frac{1}{\sqrt{2}}G^0\\  G^-\end{array}\right) \, ,
\label{eq:goldBos}
\end{eqnarray}
where $\tan\beta$ is the ratio of the two Higgs vevs, $G^\pm$ are the charged Goldstone bosons,
and $G^0$ is the neutral Goldstone boson. Since we expect that
$\tan\beta>1$, $\cos\beta$ is generally small and the couplings of the components
of the Goldstone bosons in the $H_d$ multiplet are suppressed.
On the other hand, the interactions of the components of the Goldstone bosons in the $H_u$ multiplet
can be quite large. Since the stop interacts with $H_u$, it can have a large coupling with the charged and neutral
Goldstone bosons, especially for the larger values of $\tan\beta$ considered below.

Thus, the relevant interactions of the stop with the Goldstone bosons arise from its interactions with the $H_u$ multiplet:
\begin{eqnarray}
-{\cal L}\supset  y_t (A_t H_u+\mu H_d^\dagger) \tilde Q_L \tilde t
+|y_t|^2\left(|\tilde Q_L|^2|H_u|^2+|\tilde t|^2|H_u|^2|\right) ~.
\label{eq:StopGold}
\end{eqnarray}
As we have already discussed, since $y_t> g_2$, these interactions are dominant in electroweak scattering processes for the stop.
What is less obvious is that these are also more important than the scattering processes controlled by the strong coupling, $g_3$.

\begin{figure}[t]
  \centering
  \includegraphics[width=.8\textwidth]{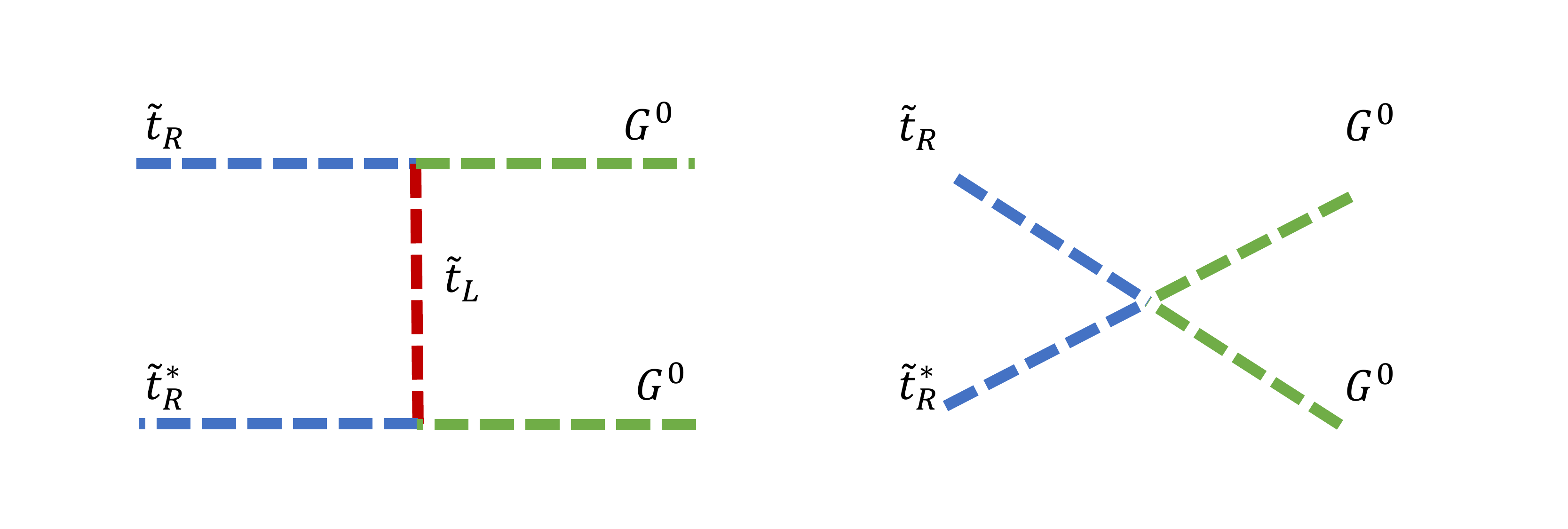}
  \caption{\it Leading-order Goldstone-boson contributions to $\tilde t_R \tilde t_R^*$ annihilation.}
  \label{fig:GoldBos}
\end{figure}

To show the significance of the scattering of stops into the longitudinal components of the $W$ and $Z$~\footnote{The significance of the
Goldstone boson mode was also discussed in \cite{Harz:2014gaa}.},
we display the leading-order contribution to these processes.
Calculating them requires the choice of a gauge. In unitary gauge,
the Goldstone bosons disappear from the theory and become the longitudinal components of the gauge bosons,
and it is difficult to see the origin of the enhancement to the scattering of the gauge bosons in this gauge,
since the Goldstone bosons are not manifest. However, in the equivalent Feynman gauge,
the Goldstone bosons {\it are} present explicitly and their contributions can quite easily
be separated from other contributions . The most important contributions of the Goldstone bosons
to the annihilation process $\tilde t_R \tilde t_R^*$ can be seen in Fig.~\ref{fig:GoldBos}~\footnote{There are also diagrams
involving the Goldstone bosons that contribute to $\tilde t_R \tilde t_R^*\to G^0 Z$, though their
contributions to the $\tilde t_R \tilde t_R^*\to ZZ$ amplitude are suppressed by a factor of $g_2$,
and can be neglected for  the purposes of this discussion. However, all contributions to
$\tilde t_R \tilde t_R^*\to ZZ$ are included in our numerical calculation.}, with analogous diagrams for the charged Goldstone boson mode.

Using the interactions of the stop and Goldstone bosons found in (\ref{eq:StopGold}),
the dominant s-wave contribution to the thermally-averaged annihilation cross sections are found to be
\begin{eqnarray}
\langle \sigma v\rangle_{\tilde t \tilde t^*\to W^+W^-}\simeq 2 \langle \sigma v\rangle_{\tilde t \tilde t^*\to ZZ}\simeq \frac{g_2^4}{128 \pi m_{\tilde t_R}^2} \left(\frac{m_t}{m_W}\right)^4 \left(\frac{\left(A_t+\mu\cot \beta\right)^2-m_{\tilde t_R}^2-m_{\tilde t_L}^2}{m_{\tilde t_R}^2+m_{\tilde t_L}^2}\right)^2+ \dots \, ,
\label{eq:ttWW}
\end{eqnarray}
where the $\dots$ represent contributions that are smaller by a factor ${\cal O}(m_W^2/m_t^2)$.
As can be seen from this expression, there are two ways in which this process is enhanced.
The first is because $m_t/m_W>1$, which is the same enhancement found in the decay $t\to bW$,
and the second is unique to scalars.  Because $A_t^2$ can be larger than $m_{\tilde t_R}^2+m_{\tilde t_L}^2$,
there is an additional possible enhancement of this annihilation process.
For $A_t^2\gg m_{\tilde t_R}^2+m_{\tilde t_L}^2$ the $\tilde t_R \tilde t_R^*$ annihilation rate is
greatly increased, and the length of the stop coannihilation strip is significantly extended.

However, $|A_t|$ cannot be increased without bound. If $|A_t|$ becomes too large,
one of the stop masses becomes tachyonic. This occurs when $A_t$ is of order $\sim m_{SUSY}^2/v$.
If such a large value of $A_t$ was allowed, the stop coannihilation trip would have no end~\footnote{This is due to the fact that the
scattering cross section in Eq. (\ref{eq:ttWW}) would scale as $m_{SUSY}^2/v^4$ instead of $1/m_{SUSY}^2$.}
However, if $A_t$ is much larger than the soft supersymmetry-breaking scalar masses,
small changes in the RG scale would lead to large changes in the soft masses. In the context of a UV-complete model,
this suggests that the mass spectrum required at the input scale is rather contrived. Even more troubling,
the mass of the SM-like Higgs boson becomes very sensitive to $A_t$ when it is much larger than the stop masses.
Indeed, as $A_t$ is increased, the SM-like Higgs boson masses is driven to zero. For these reasons,
it is expected that $A_t$ cannot be much larger than the sfermion masses~\footnote{When the Higgs mass is calculated using the code {\tt FeynHiggs~2.13.0},
our default option, it generally provides a stronger constraint than does vacuum stability.}. Even with this restriction on the size of $A_t$,
the scattering cross section in (\ref{eq:ttWW}) still gives an important boost to the stop-antistop annihilation rate.

The above restrictions constrain the amount of enhancement of the scattering cross section in (\ref{eq:ttWW}) for the CMSSM.
To maximize this enhancement, one may consider degenerate left- and right-handed stop masses.
Such a degeneracy helps because the corrections that reduce the Higgs mass are $\propto A_t^2/(m_{\tilde t_R} m_{\tilde t_L})$,
whereas the enhancement to the scattering cross section in (\ref{eq:ttWW}) is $\propto A_t^2/({m_{\tilde t_R}^2 + m_{\tilde t_L}^2})$.
The ratio of the enhancement in the scattering cross section to the reduction in the Higgs mass is therefore
$\propto (m_{\tilde t_R} m_{\tilde t_L})/(m_{\tilde t_R}^2 + m_{\tilde t_L}^2)$, which is maximized when the $m_{\tilde t_{R,L}}$
are equal. A class of models that have more degenerate stop masses are sub-GUTs,
in which the RG running of the masses is reduced. As we will see below, this increased degeneracy
indeed leads to acceptable dark matter densities with larger LSP masses.

\section{Bound-State Effects in Stop Coannihilation}

Another important effect that can lengthen the stop coannihilation strip is bound-state formation~\cite{ELO,eelo,LL}.
When dark matter froze out, at a temperature $T\sim m_{\tilde t}/25 \gg \Lambda_{QCD}$,
QCD was a relatively long-range force, and strongly-interacting particles could form bound states.
The formation rate of these states depends on the form of the long-range potential.
For a non-Abelian force, the long-range potential takes the form
\begin{eqnarray}
V(r)=-\frac{\xi}{r} \, ,
\end{eqnarray}
where $\xi$ is determined by Casimir coefficients, $C_{X_1}, C_{X_2}$, of the colour representations of the
individual particles ($X_1$ and $X_2$) forming the bound state, as well as the combined Casimir coefficient of the two particles, $C_{X_1X_2}$:
\begin{eqnarray}
\xi =\frac{1}{2} \left(C_{X_1} +C_{X_2}-C_{X_1X_2}\right)\alpha_s \, .
\end{eqnarray}
Since the gauge particle of a non-Abelian force is charged, if a gauge particle is
emitted in the formation of the bound state, the Casimir of the bound state will in general
be different from the combined Casimir of the initial-state particles.  For example,
a pair of SU(3) coloured particles in an octet configuration can transition to a bound state in a singlet representation
via the emission of a gluon.
%Since the potential for a pair of coloured particles in a singlet state is attractive,
%a bound state can form in this way.
%Because the potential in the initial configuration of the particles is repulsive,
%the energies of the particles in the thermal bath need to be large enough
%to drive the particles over this potential barrier to form the bound state with an attractive potential.
If the cross section for the formation of these bound states is large,
it alters how the constituents particles freeze out, which can be relevant when these
coloured particles are coannihilating with a dark matter candidate.

The relevance of  bound-state formation for coannihilation depends on
whether or not the bound state, ${\tilde R}$, decays more quickly than it disassociates  \cite{ELO}:
\begin{eqnarray}
\langle \sigma v\rangle_{{\tilde t} {\tilde t^*} \rightarrow {\rm SM}}
\rightarrow \langle \sigma v\rangle_{{\tilde t} {\tilde t^*} \, incl. \, \tilde{R}} \equiv \langle \sigma v\rangle _{{\tilde t} {\tilde t^*} \rightarrow {\rm SM}} + \langle \sigma v\rangle_{bsf} { {\langle  \Gamma\rangle_{\tilde R} }  \over {\langle  \Gamma\rangle_{\tilde R} + \langle \Gamma \rangle_{dis}} } \, ,
\label{eq:effglgl}
\end{eqnarray}
where $\langle \sigma v\rangle_{bsf}$, $\langle  \Gamma\rangle_{\tilde R}$,
and $\langle \Gamma \rangle_{dis}$ are the thermally averaged formation cross section, decay rate
and disassociation rate of the bound state, and $\langle \sigma v\rangle _{{\tilde t} {\tilde t^*} \rightarrow {\rm SM}}$ is the Sommerfeld-enhanced thermally-averaged cross section \cite{eoz} excluding bound-state formation. If the bound state decays much more quickly than it disassociates,
the bound state formation cross section contributes to the thermally-averaged cross section.
Because the thermally-averaged cross section is increased by this process,
the relic density is decreased for a given set of parameters and, thus,
a cosmologically-acceptable relic density can be obtained for larger sparticle masses.

In the specific case of the stop, stop-antistop pairs can form bound states through the emission of a gluon.
These bound states then decay to Standard Model particles. Since the decay rates of the bound states
are related to the scattering rates of the corresponding particles, the impact of bound-state formation will be
further enhanced by the dynamics of the Goldstone modes if decays of these bound states
through the Goldstone components of the $W/Z$ are dominant.

\section{Stop Coannihilation in the CMSSM}

In this Section we re-examine the stop coannihilation strip in the CMSSM for large $A_0$,
paying close attention to the effects of annihilations to $WW/ZZ$ and bound-state effects.
We also examine the constraints on the extent of stop coannihilation strip imposed by
the Higgs mass. However, because theoretical calculations of the Higgs mass are quite uncertain in this regime,
we first examine the stop strip independently of the Higgs mass.

\subsection{The Extent of the Stop Coannihilation Strip}

In order to understand the enhancements of the length of the stop coannihilation strip in the CMSSM due to annihilations to $WW/ZZ$
and to bound-state effects, we use the {\tt SSARD} code \cite{SSARD} to compute the particle mass spectrum and relic density.
We study the stop coannihilation strip as a function of $m_{1/2}$ for various values of $A_0$ and both signs of the
Higgsino mixing parameter $\mu$.  In each plot, the range of $m_0$ is not stated explicitly,
but is chosen such that the lighter stop mass is nearly degenerate with the LSP, which is always the lightest neutralino,
and generally the Bino.

\begin{figure}[ht!]
  \centering
  \includegraphics[width=.45\textwidth]{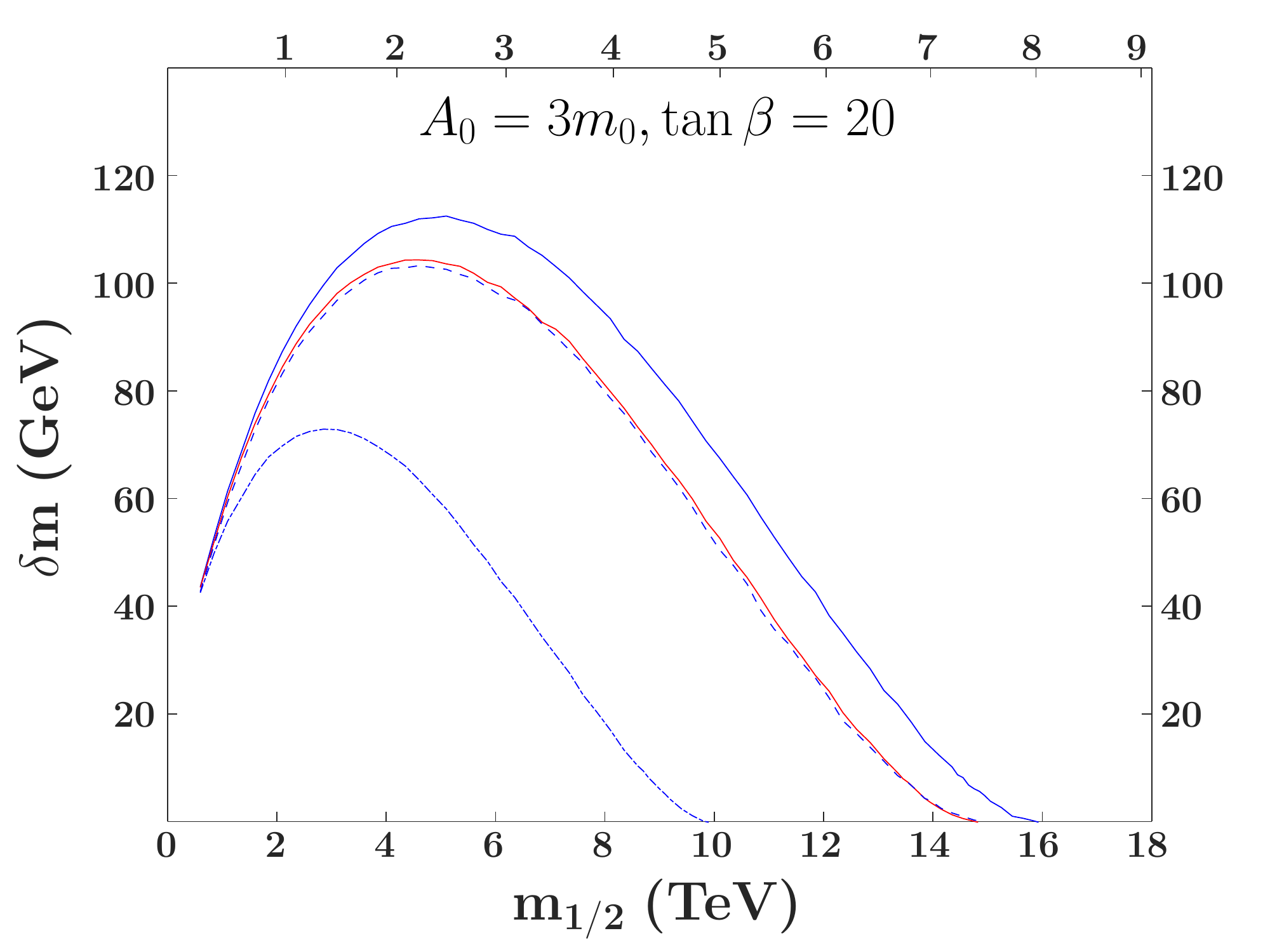} \includegraphics[width=.45\textwidth]{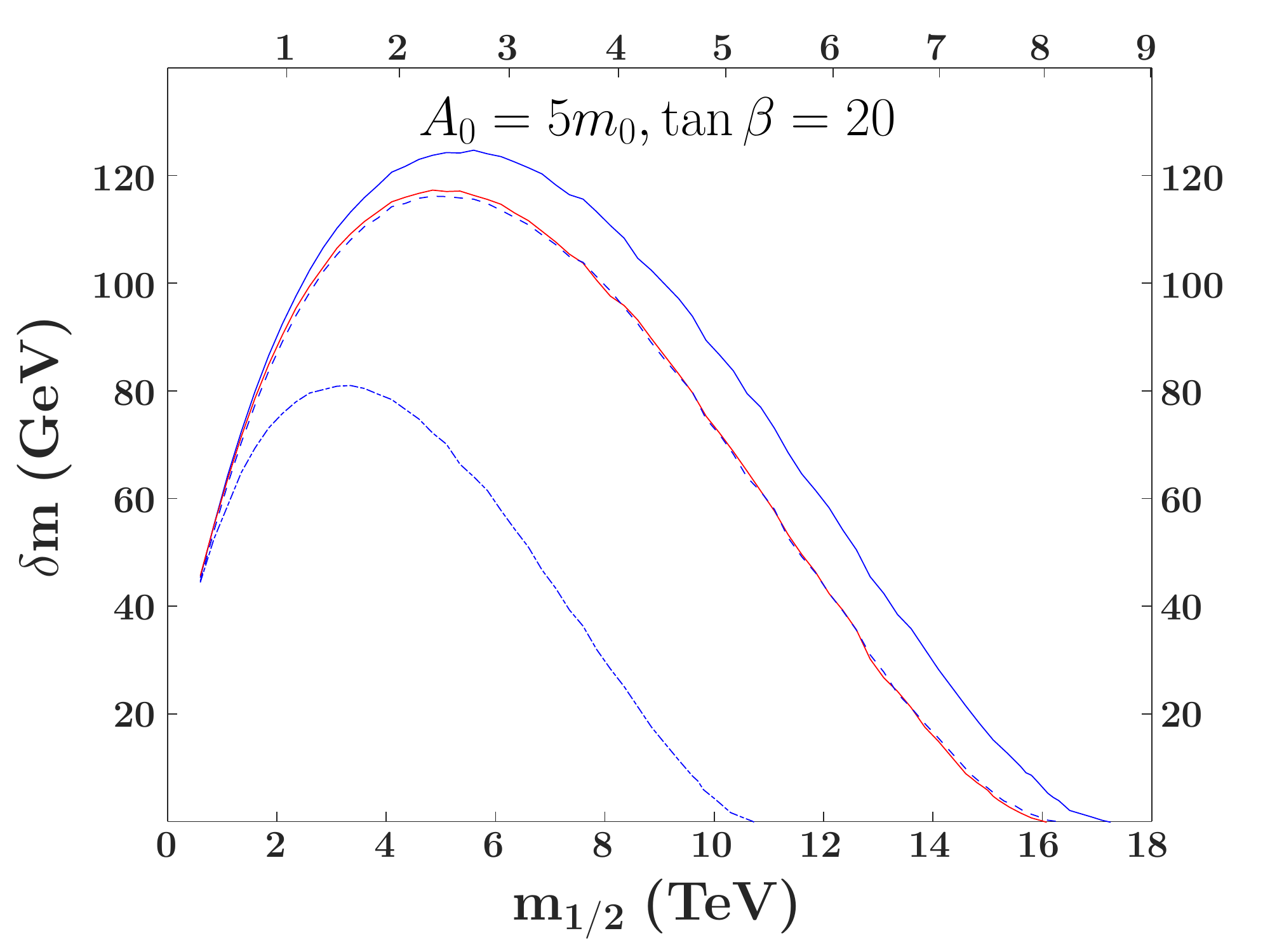} \\
  %\vspace{-3.5cm}
   {\includegraphics[width=.45\textwidth]{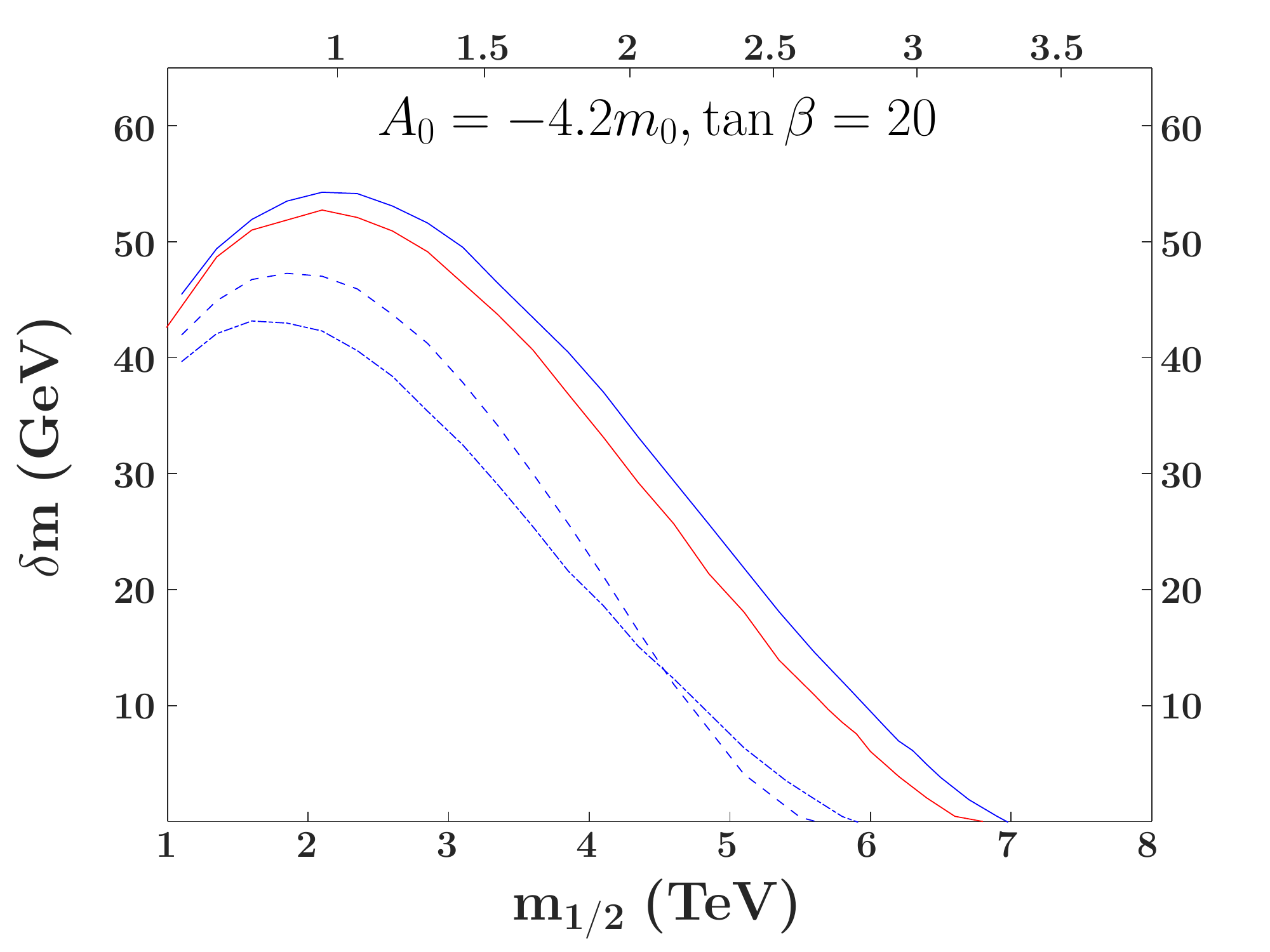} \includegraphics[width=.45\textwidth]{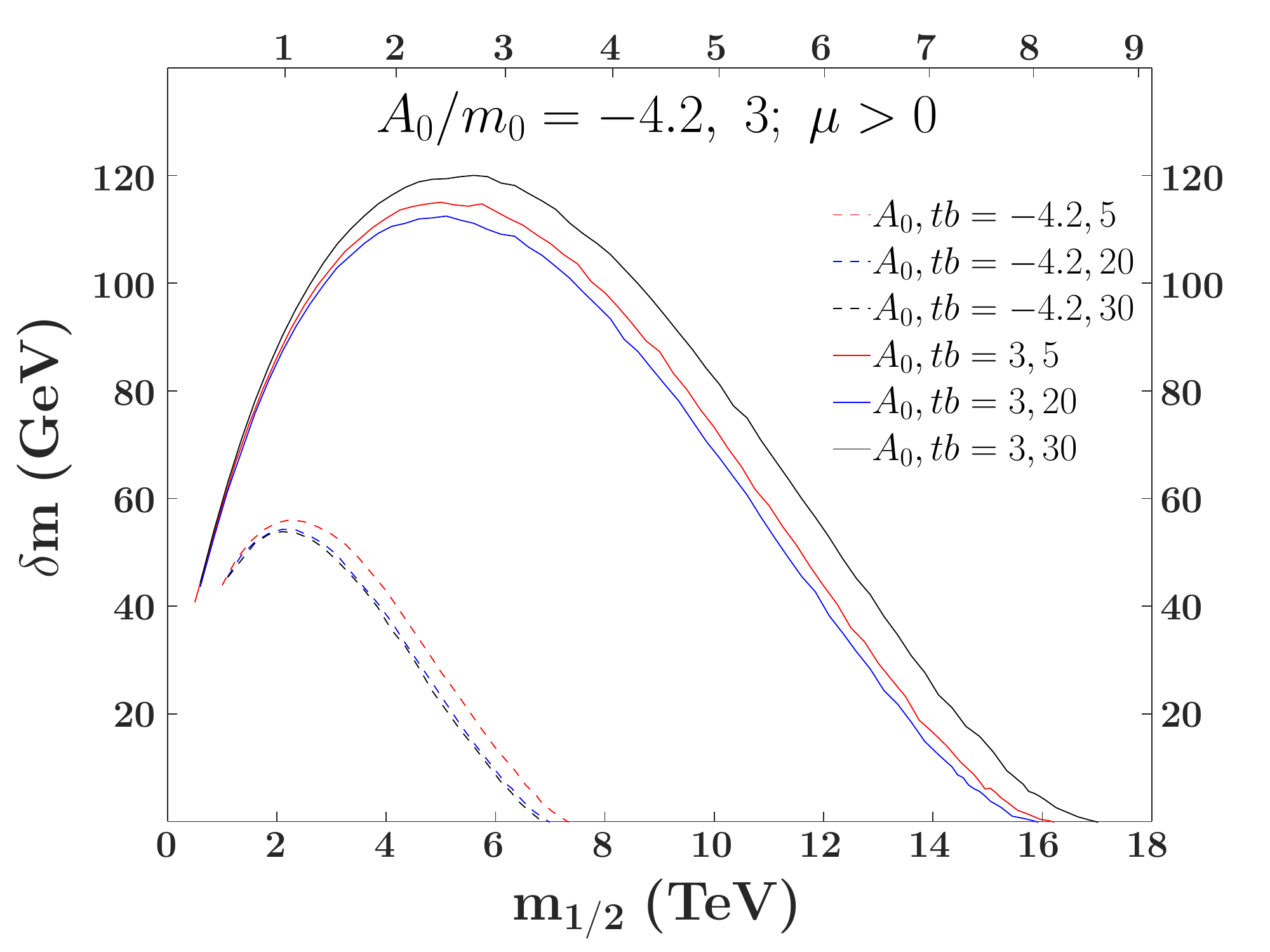}}
  \caption{\it The ${\tilde t_1} - $neutralino mass difference $\delta m$ as a function of $m_{1/2}$ along the
  stop coannihilation strips in the CMSSM for $\tan \beta = 20$ with $A_0=3m_0$ (upper left panel),
  $A_0=5m_0$ (upper right panel) and $A_0= - 4.2m_0$ (lower left panel). The solid line includes both bound states and
  scatterings to $WW/ZZ$ for $\mu>0$ (blue) and $\mu<0$ (red), the dashed line excludes only
  the bound-state effects, and the dash-dot line excludes only scatterings to $WW/ZZ$.
  The lower right panel compares results of $\tan\beta = 30$ (black), $20$ (blue) and $5$ (red)
for $A_0= 3 m_0$ (solid lines) and for $A_0=-4.2m_0$ (dashed lines) respectively.
  \label{fig:CMSSM35-42mu}}
\end{figure}

In Fig.~\ref{fig:CMSSM35-42mu} we show the mass difference $\delta m$ between the lighter stop and the LSP
(left vertical axis) that gives the correct relic density as a function of $m_{1/2}$ (lower horizontal axis)
and the corresponding values of the LSP mass, $m_\chi$ (upper horizontal axis),
for $A_0=3m_0$ (upper left panel), $A_0=5m_0$ (upper right panel)
and $A_0= -4.2 m_0$ (lower left panel), all with $\tan\beta=20$, $\mu>0$ with blue lines and $\mu<0$ with red
lines. The solid lines include both the bound-state
effect and annihilations to $WW/ZZ$. The dashed lines exclude the bound-state effect and the dash-dotted lines exclude annihilations to $WW/ZZ$.
 As is clear from these panels, the annihilations to $WW/ZZ$ are extremely important for large positive $A$-terms.
This is due to the enhancement of annihilation to the longitudinal components of the $W$ and $Z$ discussed above.
The bound-state effects, although less significant, also give an important boost to the extent of the stop coannihilation strip.
Including both the $WW/ZZ$ final states and bound-state effects, the stop coannihilation strip for $A_0=3m_0$,
$\tan\beta=20$ and $\mu>0$ extends to $m_{1/2} \sim 16$~TeV, compared to $< 10$~TeV if the $WW/ZZ$ final states
are omitted, and $< 15$~TeV if bound-state effects are omitted. The corresponding numbers for $A_0=5m_0$ are
$m_{1/2} \sim 17$~TeV, $< 11$~TeV and$\sim 16$~TeV, respectively. The corresponding maximum
values of $m_\chi$ when both these effects are included are $\sim 8$ and $\sim 8.5$~TeV, respectively.

If the sign of the $A$-term is flipped, the effect of the stop annihilations to $ZZ/WW$ is diminished,
as can be see in the lower left panel of Fig.~\ref{fig:CMSSM35-42mu}, where we again plot the mass difference $\delta m$
as a function of $m_{1/2}$ and $m_\chi$ for $A_0 = - 4.2 m_0$, $\tan\beta=20$ and $\mu>0$.
This reduction in the effect of the scatterings to $WW/ZZ$ is due to the
RG running of the trilinear coupling $A$. For $A_0<0$, the gaugino and trilinear contribution to the trilinear beta function drives it towards zero.
This leads to a more significant reduction in the $A$-term as it runs towards lower scales.
With the $A$-term much smaller at the SUSY scale, the coupling of the Goldstone boson to the stop is diminished,
and the scatterings to the longitudinal components of $WW/ZZ$ are suppressed. The dominant stop-antistop annihilation channel is then
$\tilde t \tilde t^* \to gg$.  Since the annihilations to $WW/ZZ$ are suppressed, the relic density tends to be much larger for $A_0<0$ for a
comparable stop and LSP mass splitting, shortening the extent of the stop strip. The bound state effects for $A_0 = - 4.2 m_0$,
on the other hand, are more significant than for the positive values of $A_0$
studied previously~\footnote{This is because the scattering rate in this regime is dominated by QCD interactions.
Since bound state formation is also governed by QCD, it roughly doubles the annihilation cross section.}.
The length of the stop coannihilation strip for $A_0 = - 4.2 m_0$
extends to $m_{1/2} \sim 7$~TeV when both the $WW/ZZ$ final states and bound-state effects are included, compared
to $m_{1/2} < 6$~TeV if either of these are omitted, corresponding to $m_\chi \lesssim 3.5$~TeV.

A direct comparison between the lengths of the stop coannihilation strips for
the different values of $A_0$ and different signs of $\mu$ when $\tan \beta = 20$ shows that the strip is shorter for
smaller values of $A_0$, particularly for $A_0 < 0$.
As can be seen in Eq. (\ref{eq:ttWW}), $\mu$ also plays a role in the annihilations to $WW/ZZ$.
The interaction of the stops with the Goldstone mode through $\mu$ is suppressed by $\cot\beta$,
since this contribution stems from an interaction with $H_d$ instead of $H_u$.  However,
even with this $\cot\beta$ suppression, it still plays a role. Comparing the results for $\mu>0$ (blue lines) and $\mu<0$ (red lines) in the upper panels and the lower left panel of
Fig.~\ref{fig:CMSSM35-42mu}, we see that the strip for $\mu < 0$ is $\sim 1$~TeV shorter than for $\mu > 0$ when
$A_0> 0$, with a smaller reduction for $A_0 = - 4.2 m_0$~\footnote{This is due to the fact that this process is dominated by QCD processes
for $A_0<0$.}. The corresponding reductions in the
maximum values of $m_\chi$ are $\lesssim 0.5$~TeV.

Some comparisons of the lengths of the coannihilation strips for different values of $\tan \beta$
are shown in the lower right panel of Fig.~\ref{fig:CMSSM35-42mu}. Those for $A_0 = 3 m_0$ are shown as solid lines,
and those for $A_0=-4.2m_0$ are shown as dashed lines. The range in the extent of the stop coannihilation strip as a function $\tan\beta$ is of order $\Delta m_{1/2} \sim 1$~TeV
for $A_0= 3m_0$, namely between $\sim 16$ and 17~TeV,
corresponding to $m_\chi \lesssim 8.5$~TeV. For $A_0=-4.2m_0$, the range in the extent of the stop strip is considerably less,
and for all values of $\tan\beta$ considered here the strip terminates at $m_{1/2}\sim 7$ TeV,
corresponding to $m_\chi \lesssim 3.5$~TeV.
We see that the strips are longest for $\tan \beta =30$ (black lines) and $\tan \beta =5$ (red lines) when $A_0/m_0=3$ and $-4.2$ respectively.  For $\tan\beta=5$, $\cot \beta$ is large enough that the $\mu$ contribution in Eq. (\ref{eq:ttWW}) is important and extends the stop coannihilation strip a small but noticeable amount for both signs~\footnote{For $A_0<0$, we have $|\mu|\cot\beta> |A|$ which leads to an enhancement of the $\tilde t\tilde t^* \to WW$ scattering.}  of $A_0$. If $\tan\beta$ is further increased  to $40$, the LSP becomes a stau or a stop for both $A_0/m_0=3$ and $-4.2$.

\subsection{The Higgs Mass along the Stop Coannihilation Strip}

\begin{figure}[ht!]
  \centering
  \includegraphics[width=.45\textwidth]{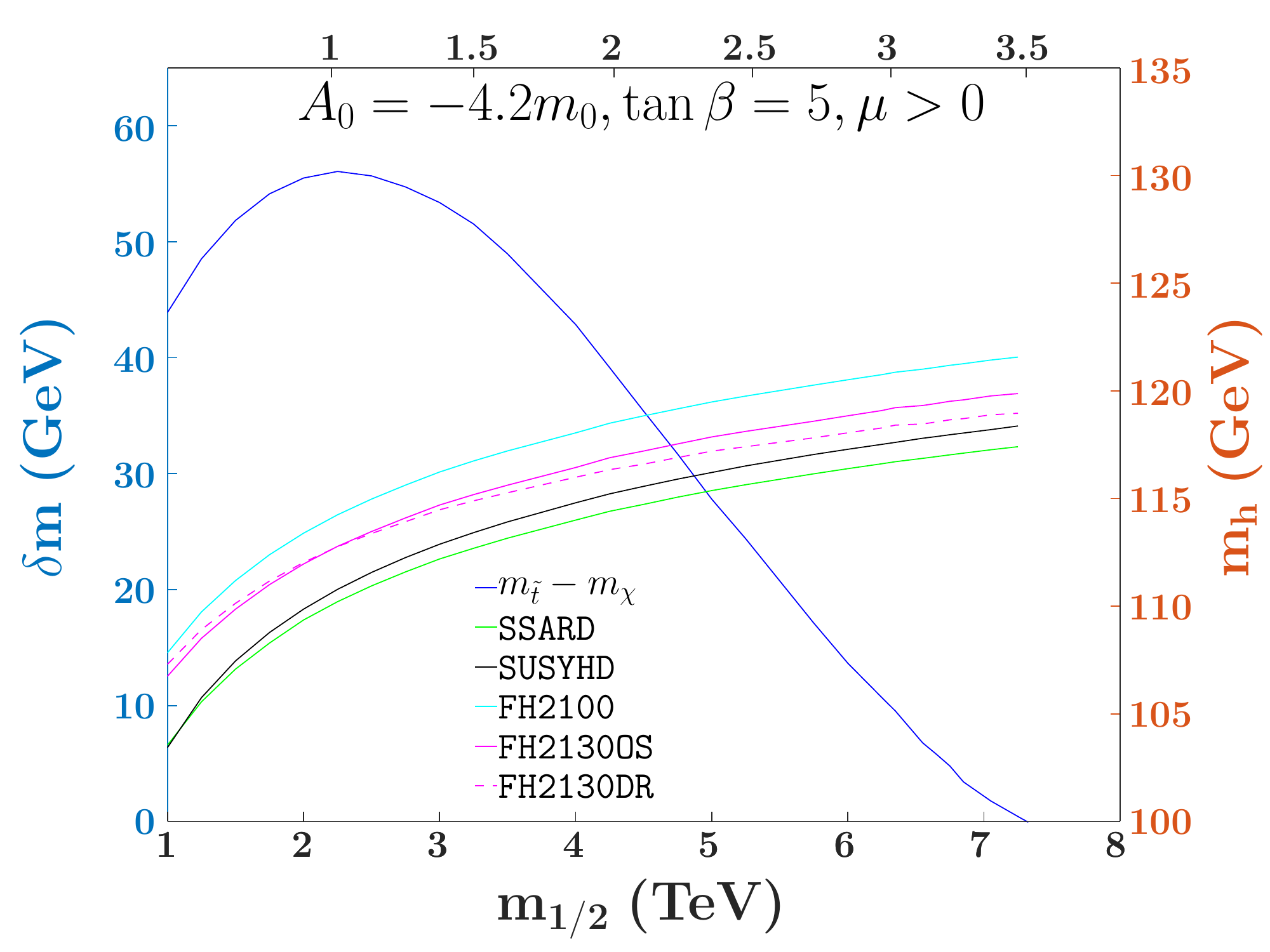}
  \includegraphics[width=.45\textwidth]{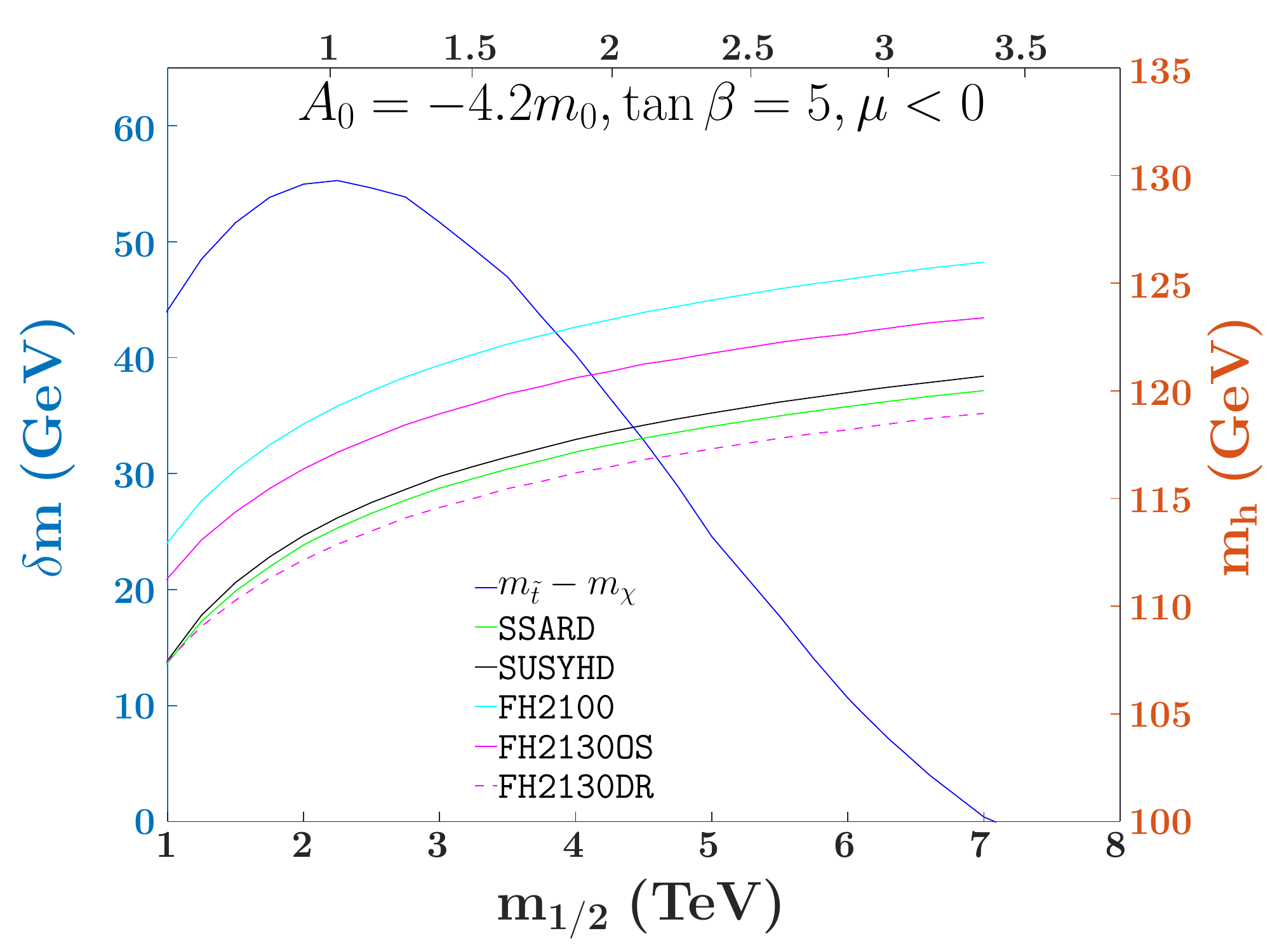}
  \includegraphics[width=.45\textwidth]{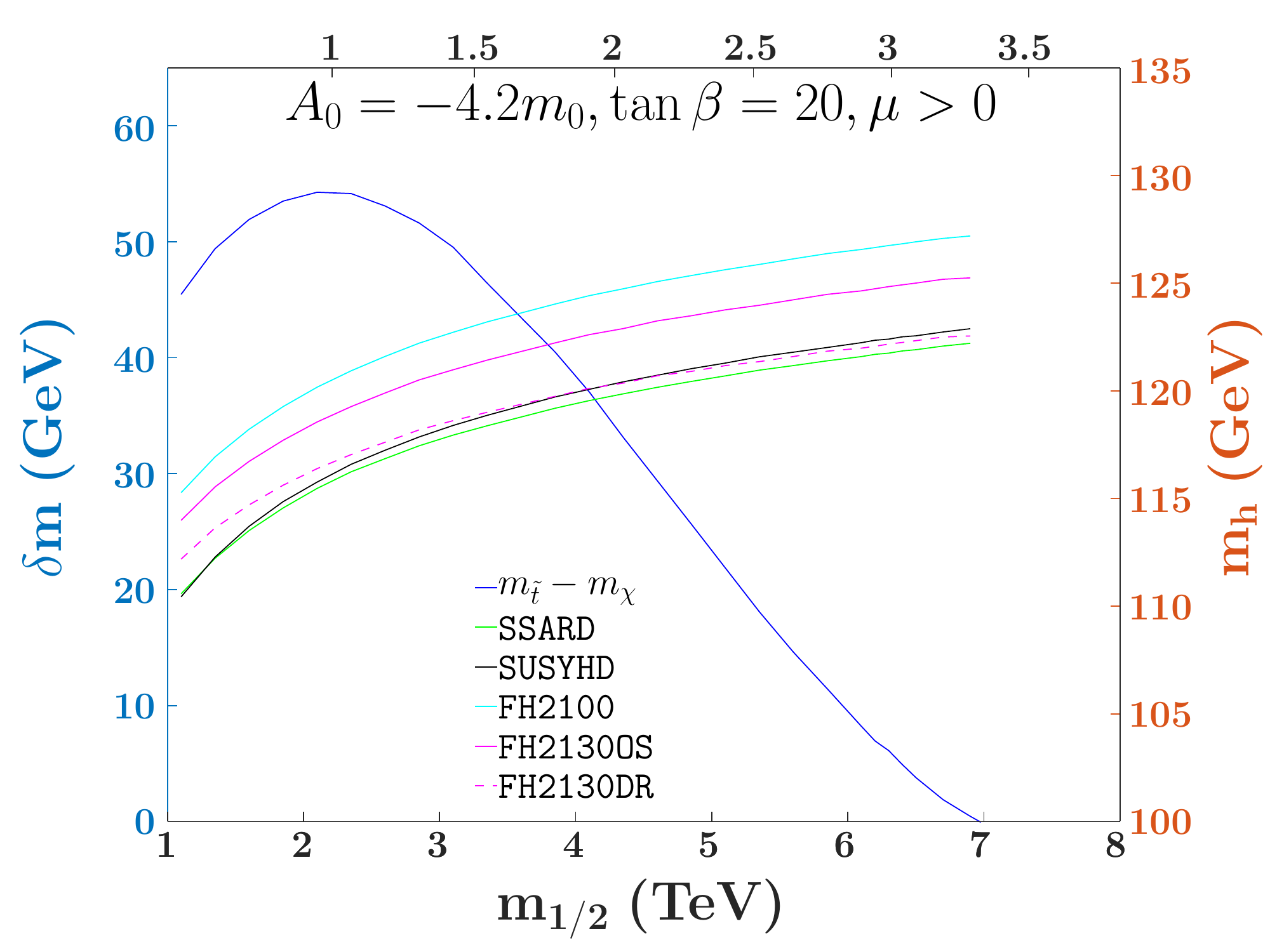}
  \includegraphics[width=.45\textwidth]{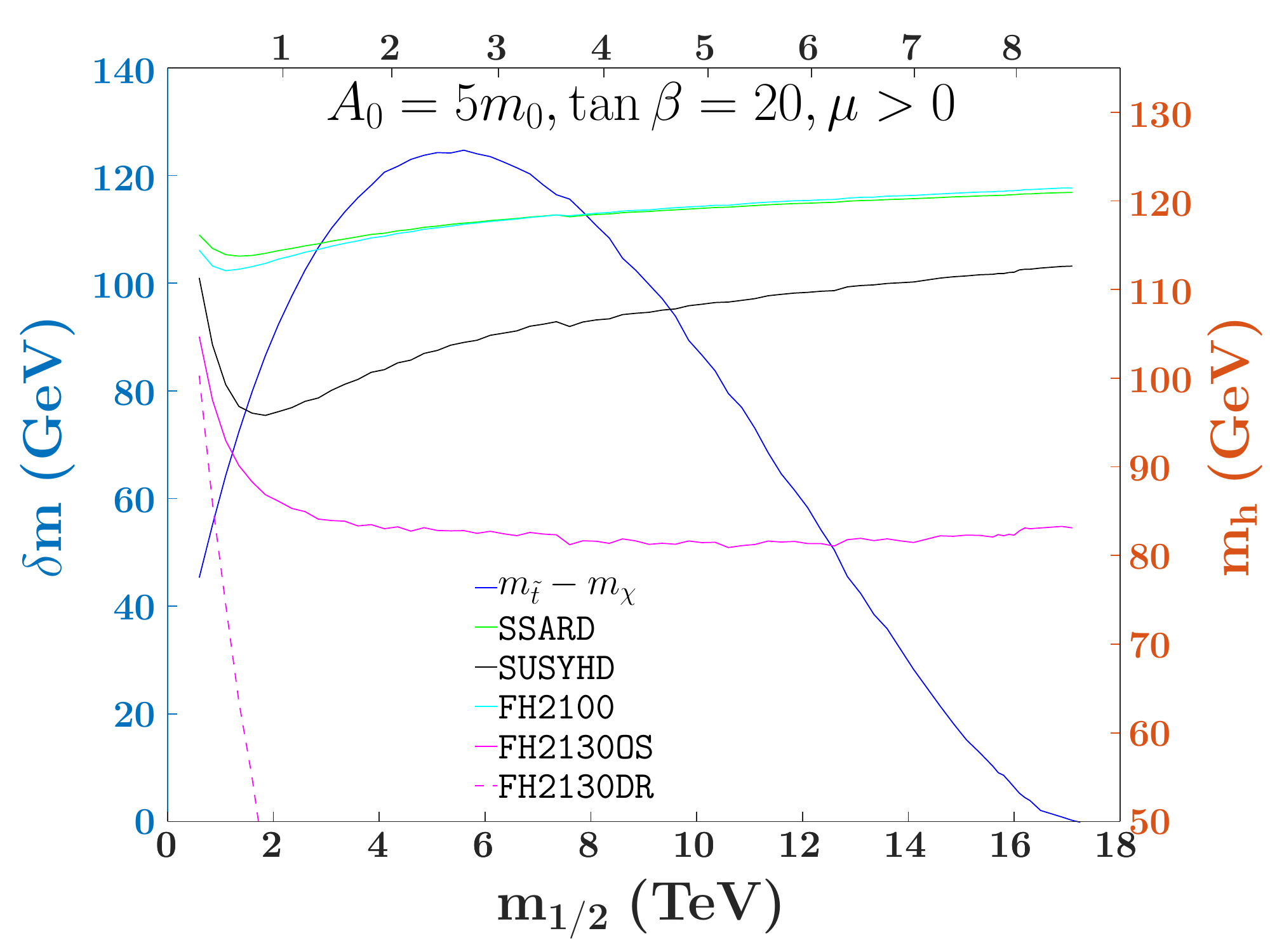}
\includegraphics[width=.45\textwidth]{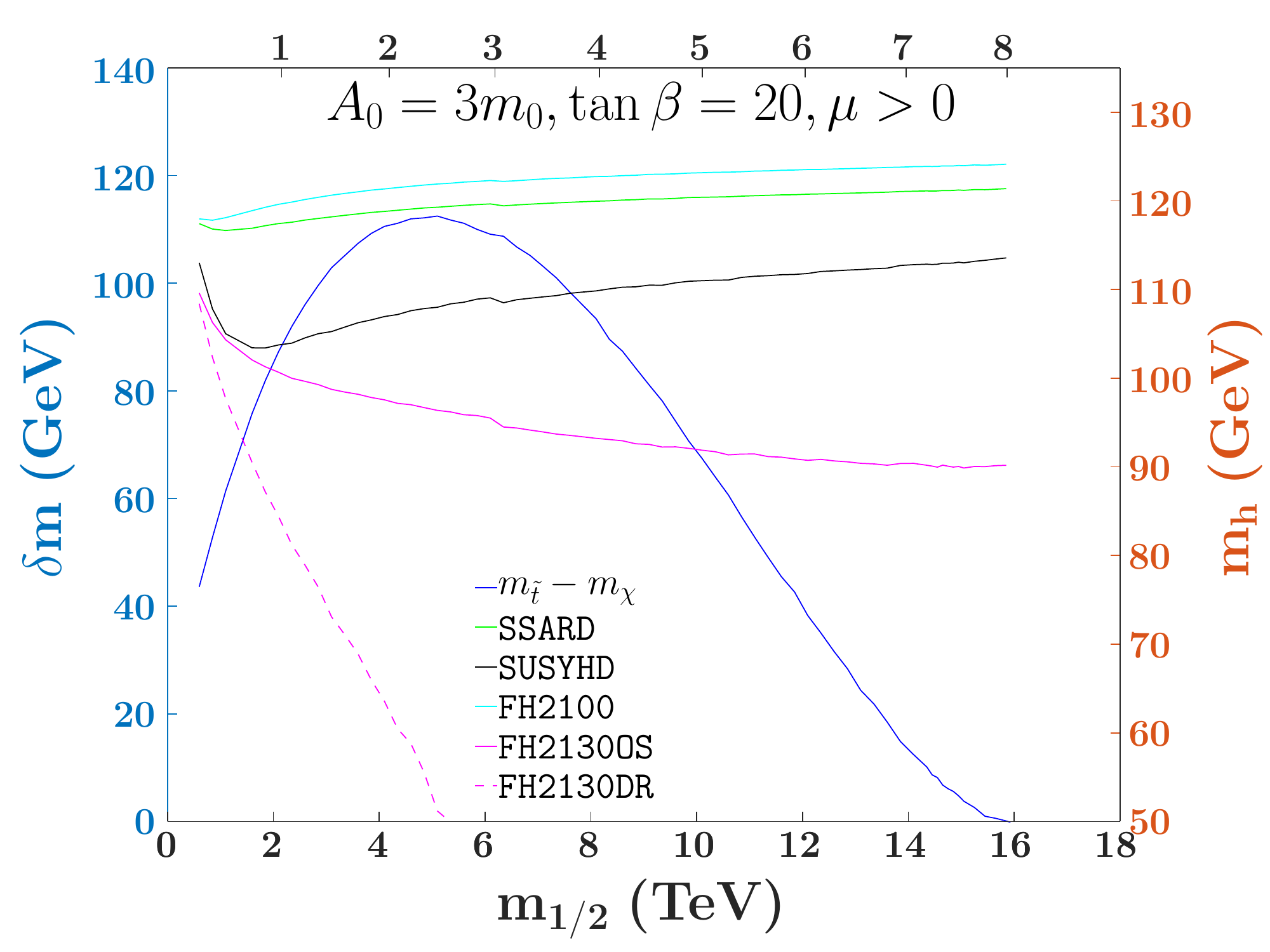}
\includegraphics[width=.45\textwidth]{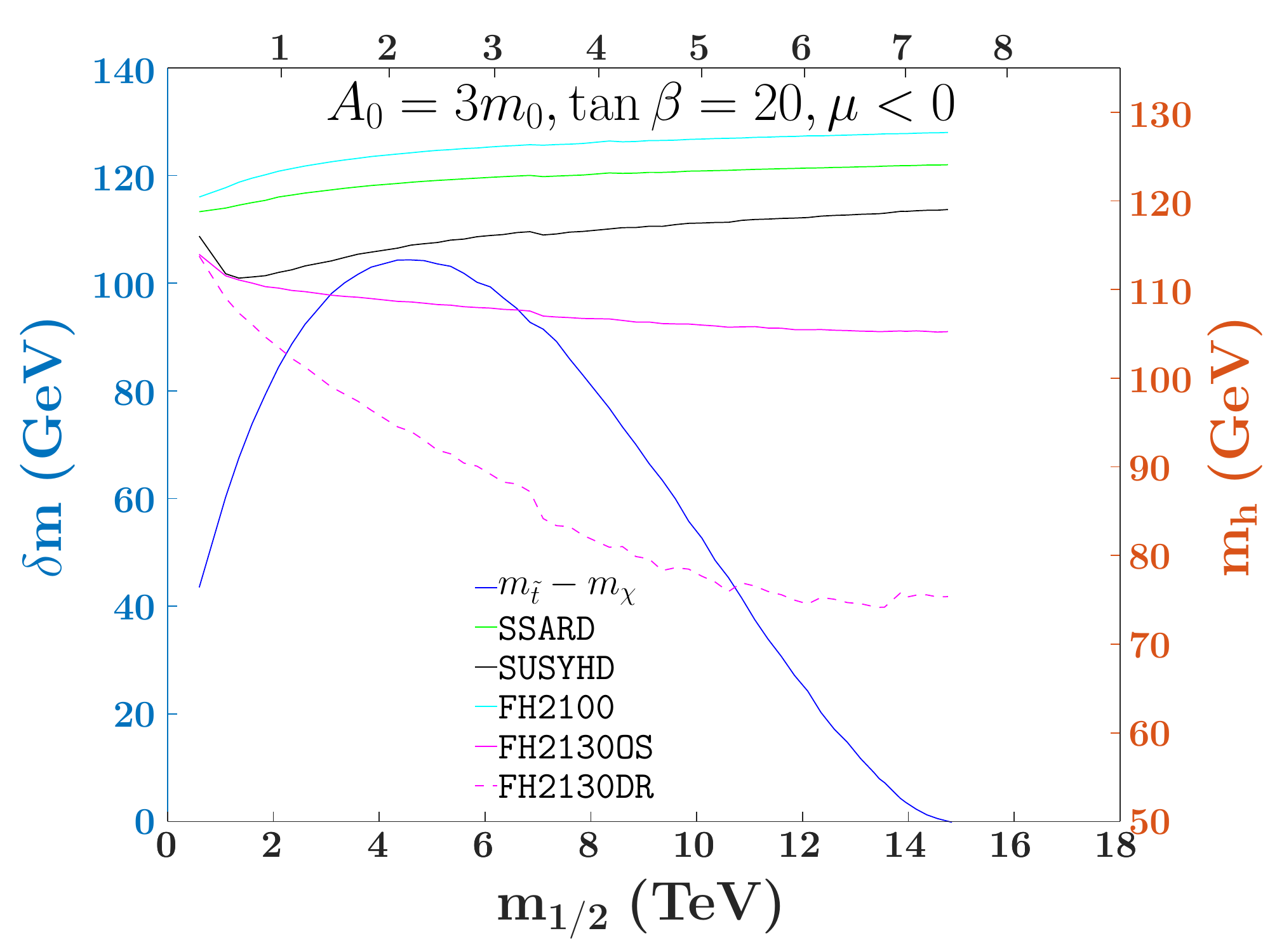}
\caption{\it The stop coannihilation strips for $A_0=-4.2m_0$, $\tan \beta =5$ and both positive or negative $\mu$
(upper panels), for $A_0=-4.2m_0$, $\tan \beta =20$ and $\mu > 0$ (middle left panel),
$A_0= 5 m_0$, $\tan \beta =20$ and $\mu > 0$ (middle right panel) and $A_0=3m_0$, $\tan \beta =20$ and both positive or negative $\mu$ (lower panels) with the Higgs mass calculated using different codes: {\tt FeynHiggs~2.10.0} (cyan lines) and {\tt 2.13.0}
(purple lines), {\tt SSARD} (green lines) and {\tt SUSYHD} (black lines).} \label{fig:3}
\end{figure}

\begin{figure}[ht!]
  \centering
  \includegraphics[width=.45\textwidth]{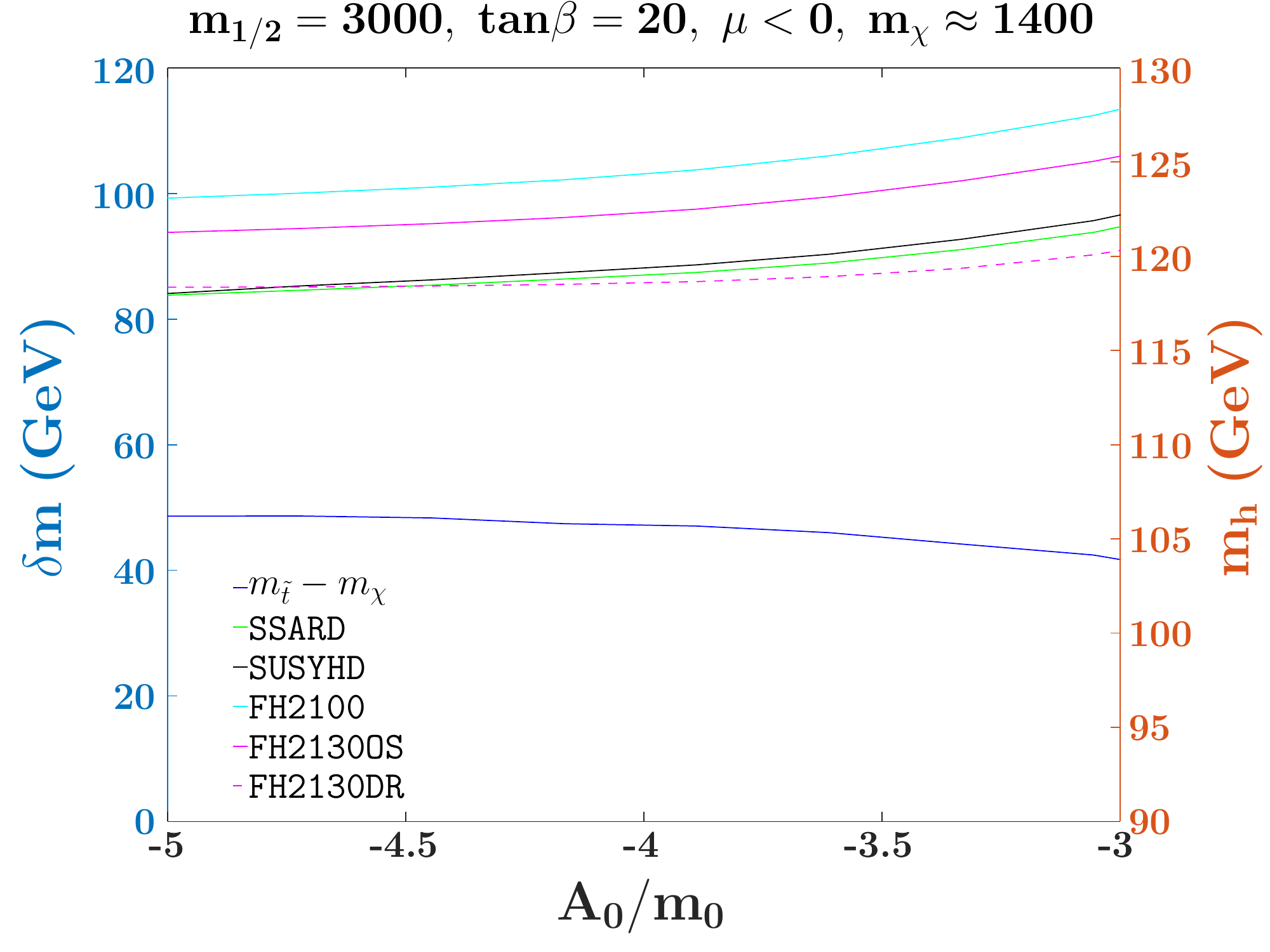}
  \includegraphics[width=.45\textwidth]{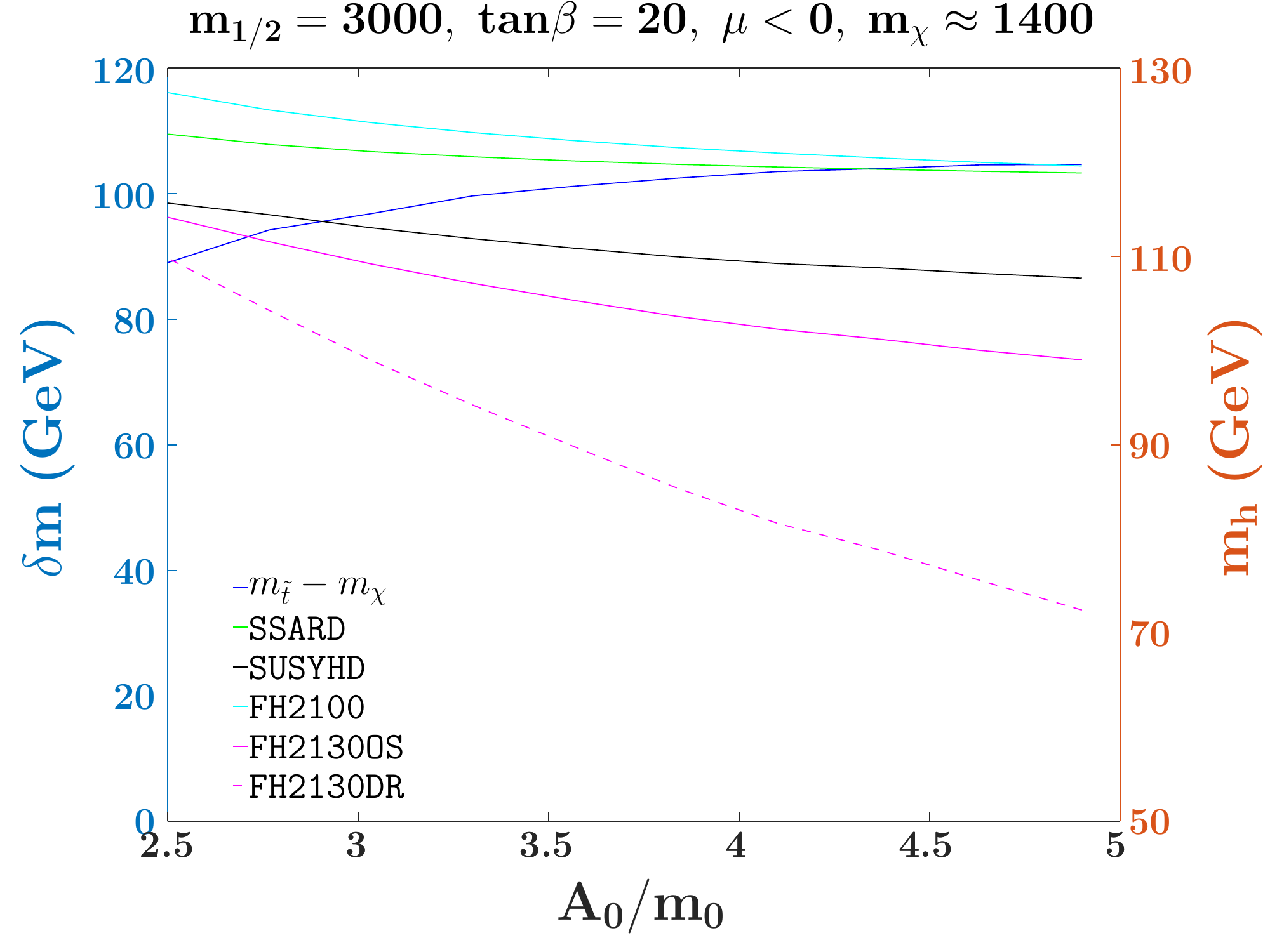}\\
\includegraphics[width=.45\textwidth]{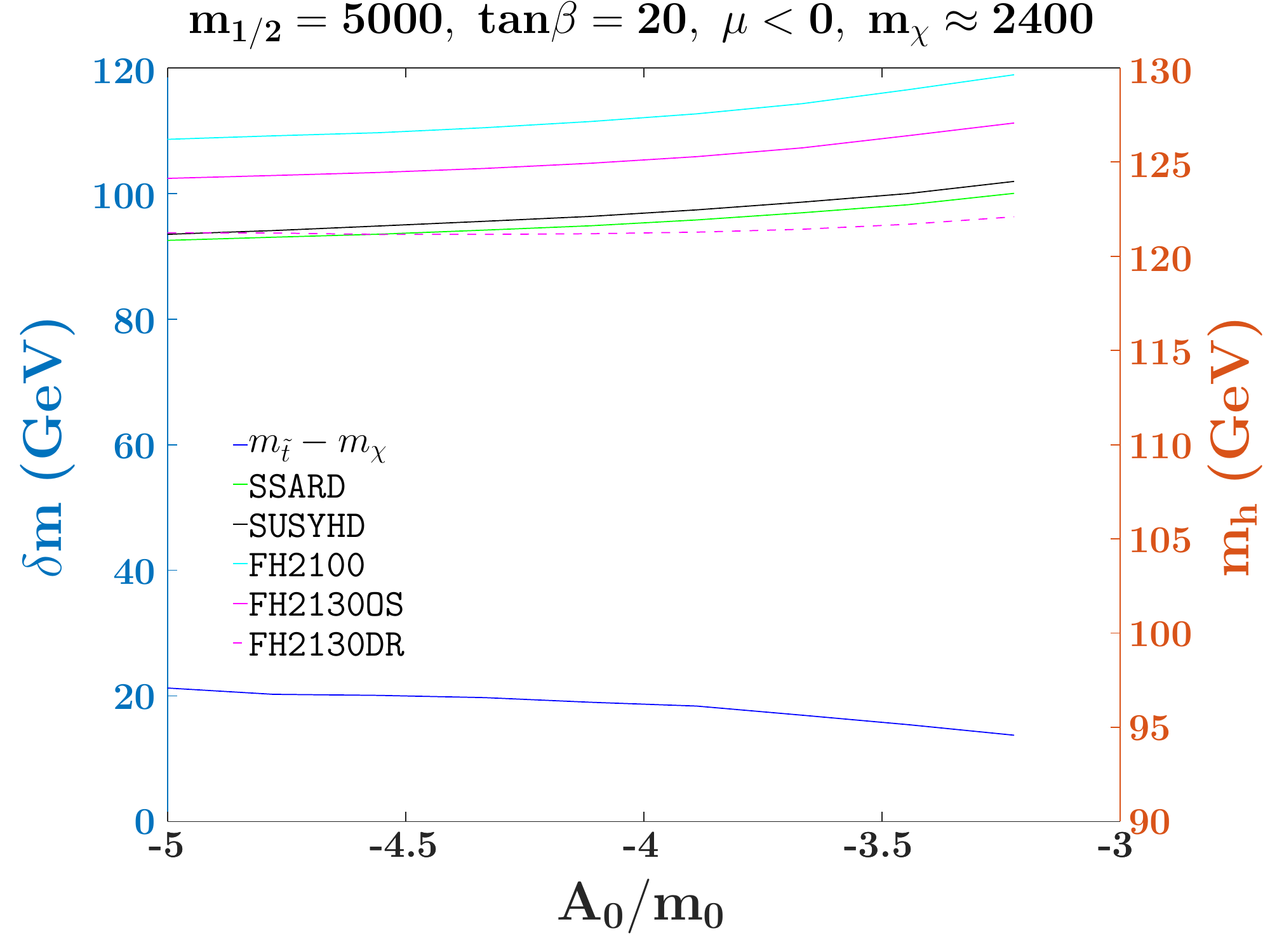}
  \includegraphics[width=.45\textwidth]{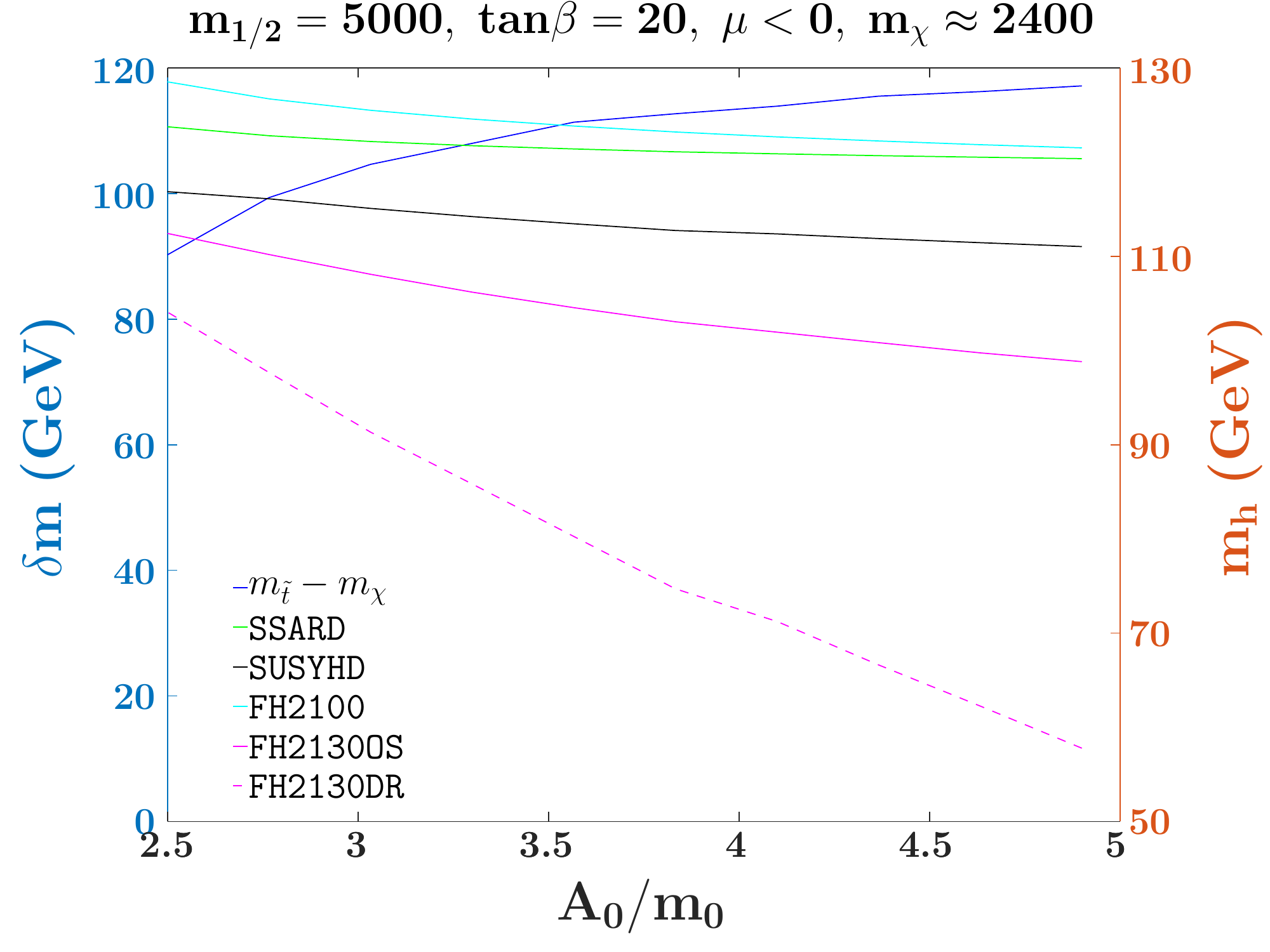}
\caption{\it Examination of Higgs mass for large $A$-terms along the stop coannihilation strip.}
\label{fig:4}
\end{figure}

We now examine the constraints on the allowable extent of the stop coannihilation strip
that are potentially imposed by the Higgs mass, comparing the results obtained using different codes for
calculating $M_h$ in the MSSM. The codes we consider in Figs.~\ref{fig:3} and \ref{fig:4} are
{\tt FeynHiggs~2.10.0} (cyan lines) and {\tt 2.13.0}
(purple lines), {\tt SSARD} (green lines) and {\tt SUSYHD} (black lines)~\footnote{The {\tt SSARD} calculation of $M_h$,
is heavily based on the works in ref. \cite{mhsplit} and is expected to be reasonably accurate only when
the SUSY mass scales are $\gtrsim$ several TeV.}. {\tt FeynHiggs~2.13.0} provides the choice of inputting parameters using either the on-shell
(OS) scheme or the dimensional reduction (DR) scheme. We also compare in the figures
the results obtained with these two different set of inputs, by converting the relevant parameters generated by the {\tt SSARD} code to
the OS scheme (purple solid lines) or DR scheme (purple dashed lines).

The two upper panels of Fig.~\ref{fig:3} show stop coannihilation strips for $A_0=-4.2m_0$,
$\tan \beta =5$ and $\mu > 0$ (left panel) or $\mu < 0$ (right panel), with the values of $m_0$ chosen to obtain the correct cosmological dark matter density, as
calculated including $WW/ZZ$ final states and bound-state effects.
The values of $\delta m = m_{\tilde t_i} - m_\chi$ are shown as solid blue lines to be read on the left vertical axis.
In these cases the values of $M_h$ for identical inputs, to be read on the right vertical axis, have a spread $\lesssim 5$~GeV,
which does not vary significantly along the strip, but is slightly larger for $\mu < 0$ than for $\mu > 0$.  Similar results for $A_0=-4.2m_0$,
$\tan \beta =20$ and $\mu > 0$ are shown in the middle left panel of the same figure, with $M_h$ higher than the $\tan \beta =5$ cases.
Since {\tt FeynHiggs~2.13.0} supersedes version {\tt 2.10.0} and it includes effects which are not part of
other codes, we take it with on-shell input masses as our default in the following sections, with an uncertainty that we estimate to be at least $3$~GeV.

In the case $A_0=-4.2m_0$, $\mu > 0$ and $\tan \beta =5$ (upper left panel of Fig.~\ref{fig:3}), the calculated value of $M_h$
is always less than the experimental value, even taking the theoretical uncertainty into account. On the other hand,
in the case $A_0=-4.2m_0$, $\mu > 0$ and $\tan \beta =5$ or $\tan \beta =20$ (upper right  and middle left panels of Fig.~\ref{fig:3}), the calculated value of $M_h$
is compatible with the measured value all the way from $m_{1/2} \sim 3$~TeV
to the end of the stop coannihilation strip at $m_{1/2} \sim 7$~TeV. Although the Higgs mass is stable,
we note that the extent of the stop strip is restricted because $A_0<0$.

For the middle right and bottom two panels of Fig.~\ref{fig:3}, we consider $A_0>0$, and the stop masses are much more split and the $A$-terms at the SUSY scale are much larger. This extreme set of soft supersymmetry-breaking parameters makes it rather difficult to calculate the Higgs mass. Since the splitting of the stop masses along the stop coannihilation strip becomes even more extreme for larger $m_{1/2}$, the variance in the Higgs mass calculators make it unclear what really is the endpoint of the stop coannihilation strip. For example, in the middle right panel, we take $A_0=5m_0$, $\tan\beta=20$ and $\mu>0$ and for the bottom left we take instead $A_0=3m_0$. In these panels, the Higgs mass using {\tt FeynHiggs~2.13.0} DR goes to zero for $m_{1/2}=5000$ GeV ($m_{1/2}=2000$ GeV) for $A_0=5m_0$ ($A_0=3m_0$). Ignoring this, the spread in the Higgs mass is still 40 GeV(35 GeV) for $A_0=5m_0$ ($A_0=3m_0$). The situation is slightly better
when $\mu<0$, where the {\tt FeynHiggs~2.13.0} DR calculation does not drop to zero. In this case, the spread including the {\tt FeynHiggs~2.13.0} DR calculation is larger than 45 GeV and without it is of order 25 GeV.  Although this case is better, it is still far too inaccurate to derive any meaningful constraints on the extent of the stop coannihilation strip.

In Fig. \ref{fig:4}, we examine the Higgs mass calculators along the stop strip as a function of the input trilinear coupling $A_0$ for $\mu<0$ and $\tan\beta=20$. In the upper (lower) panel we set $m_{1/2}=3000$ GeV ($m_{1/2}=5000$ GeV). For the two left panels, the $A_0<0$ portion is shown.  In this regime, the Higgs mass calculators agree reasonably well with each other over the entire range of $A_0$ plotted, with the variation in the Higgs mass being less than 10 GeV.  The difference in the stop and LSP mass, $\delta m$, which is needed to give an acceptable dark matter relic density is of order 40 GeV (20 GeV) for the upper left (bottom left) panel. In the right two panels, we show the $A_0>0$ parameter space where the Higgs mass calculators do not agree. The variation in the Higgs mass calculators begins at $20$ GeV ($30$ GeV) for $A_0=2.5 m_0$ and increases to $50$ GeV ($60$ GeV) for $A_0=5 m_0$ for the upper right (bottom right) panel.

\begin{figure}[ht!]
\centering
 \begin{minipage}[b]{0.45\textwidth}
    \includegraphics[width=\textwidth ]{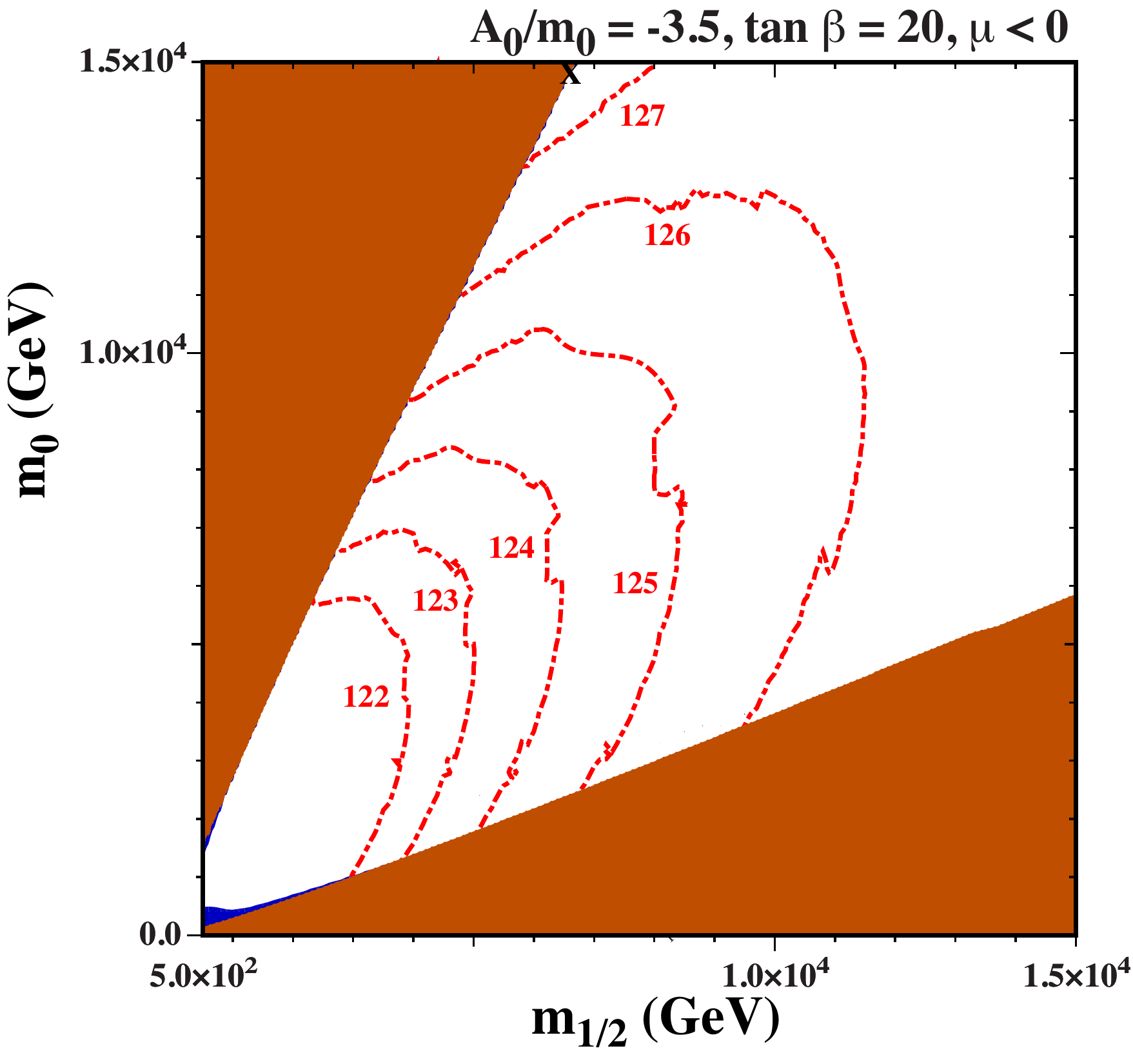}
 \end{minipage}
 \hspace{0.5cm}
  \begin{minipage}[b]{0.45\textwidth}
     \includegraphics[width= \textwidth ]{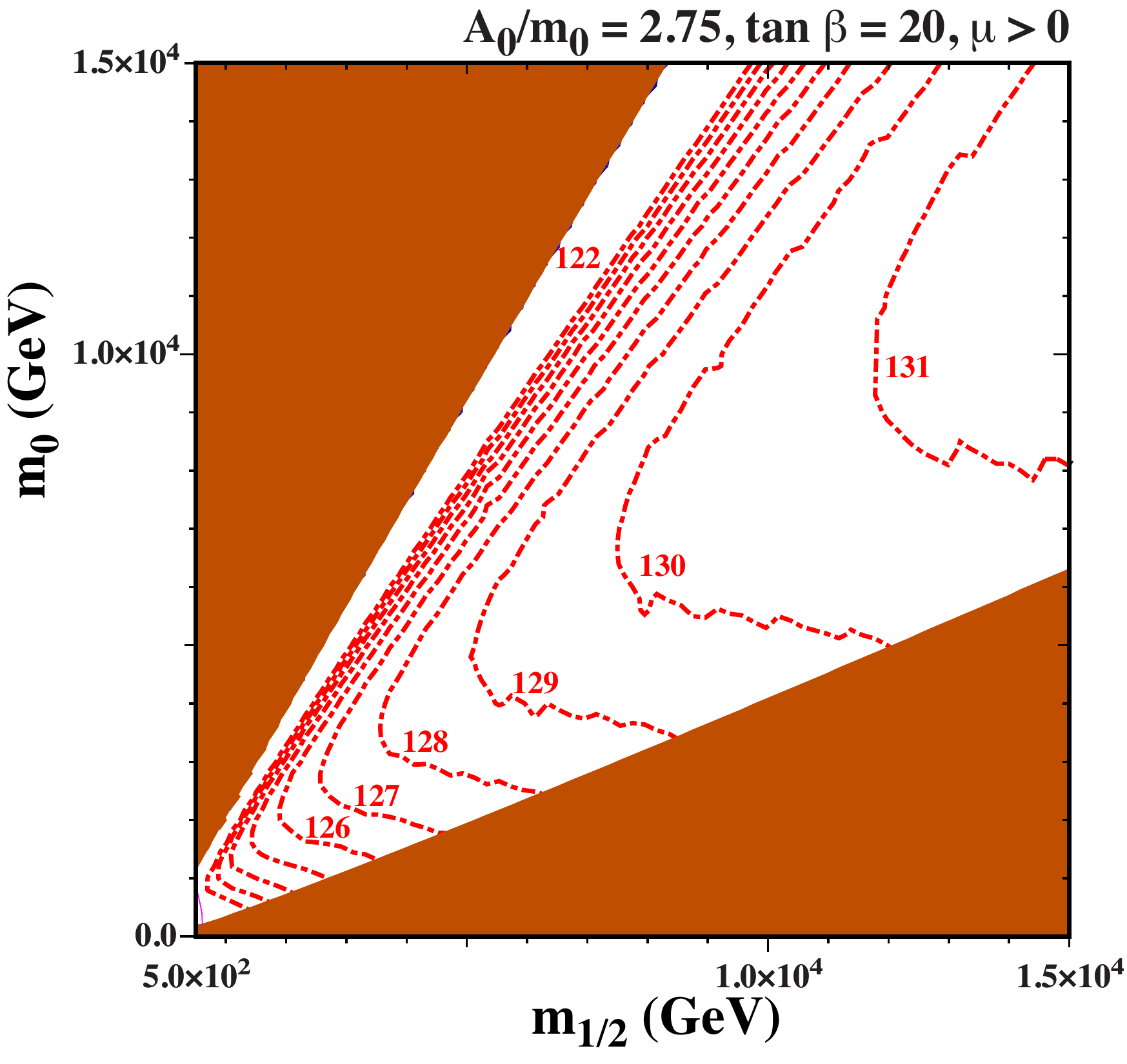}
 \end{minipage}
\caption{\it The $(m_{1/2},m_0)$ planes in the CMSSM with $\tan \beta = 20$. In the left panel,  
$A_0 = -3.5 m_0$ and $\mu < 0$, whereas in the right panel
$A_0=2.75 m_0$ and $\mu>0$. Strips with the allowed cosmological LSP density are
shaded dark blue (enhanced so that $0.01 < \Omega h^2 < 2.0$
though they are still essentially invisible). The endpoint of the strip in the left panel is marked with an X,
but is beyond the range of the right panel.
Regions where the LSP is charged are shaded brick red
and contours of $M_h$ are indicated by red dot-dashed lines.}
\label{fig:cmssmplanes}
\end{figure}

For completeness, we show two examples of ($m_{1/2}, m_0$) planes in the CMSSM with $\tan \beta = 20$ in Fig. \ref{fig:cmssmplanes}.
In the left panel, we have chosen $A_0 = -3.5 m_0$ and $\mu < 0$, and in the right panel, $A_0 = 2.75$ and $\mu > 0$.
In both panels, the brick-shaded regions are where the LSP is charged, and are for that reason excluded by cosmology.
The brick-shaded region in the lower right corner corresponds to a stau LSP,
while that in the upper left corner corresponds to a stop LSP.
 For a given value of $m_0$, the stop mass
varies very rapidly with $m_{1/2}$ making the stop coannihilation strip (shown in blue) extremely thin (essentially invisible)
even when extending the range on the cosmological density to $0.01 < \Omega h^2 < 2$.
In the left panel, the strip ends at the point marked with an X (near $(m_{1/2},m_0) = (6.6,14.8)$ TeV).
In the right panel, the stop strip extends beyond the range shown.  However, the coordinates of the end point can be surmised by examining
Fig. \ref{fig:CompCMSSMSubGut}. The most noticeable difference between the two panels
is the value of the Higgs mass across the plane: contours of the Higgs mass are shown by the red dot-dashed curves.
Consistent with the discussion above, the Higgs mass along
the strip is reasonable only when $A_0 < 0$ and $\mu < 0$. A related discussion on similar planes can be found in
\cite{eemno}.

\section{The Stop Coannihilation Strip in Sub-GUT Models}

We now discuss the stop coannihilation strip in a variant of the CMSSM in which the soft supersymmetry-breaking
sparticle masses are assumed to be universal at some renormalization scale $\Min < \MGUT$, as in `mirage unification' \cite{mirage}
and other sub-GUT models \cite{subGUT}. As was commented above, one can anticipate that the stop coannihilation strip may extend to larger LSP
masses than in the CMSSM, because the renormalization-group running of the input parameters
over a smaller range of scales allows the two stop masses
$m_{\tilde t_1}$ and $m_{\tilde t_2}$ to be more similar than in the CMSSM.  Since the Higgs squared
mass depends on $A_t^2/{m_{\tilde t_1}m_{\tilde t_2}}$ while the length of the coannihilation strip
depends approximately on $A_t^2/({m_{\tilde t_1}^2+m_{\tilde t_2}^2)}$, this enables the length of
the coannihilation strip to be maximized while retaining a value of the Higgs mass that is consistent with
experiment.

\begin{figure}[ht!]
\centering
 \begin{minipage}[b]{0.4\textwidth}
    \includegraphics[width=1.02\textwidth ]{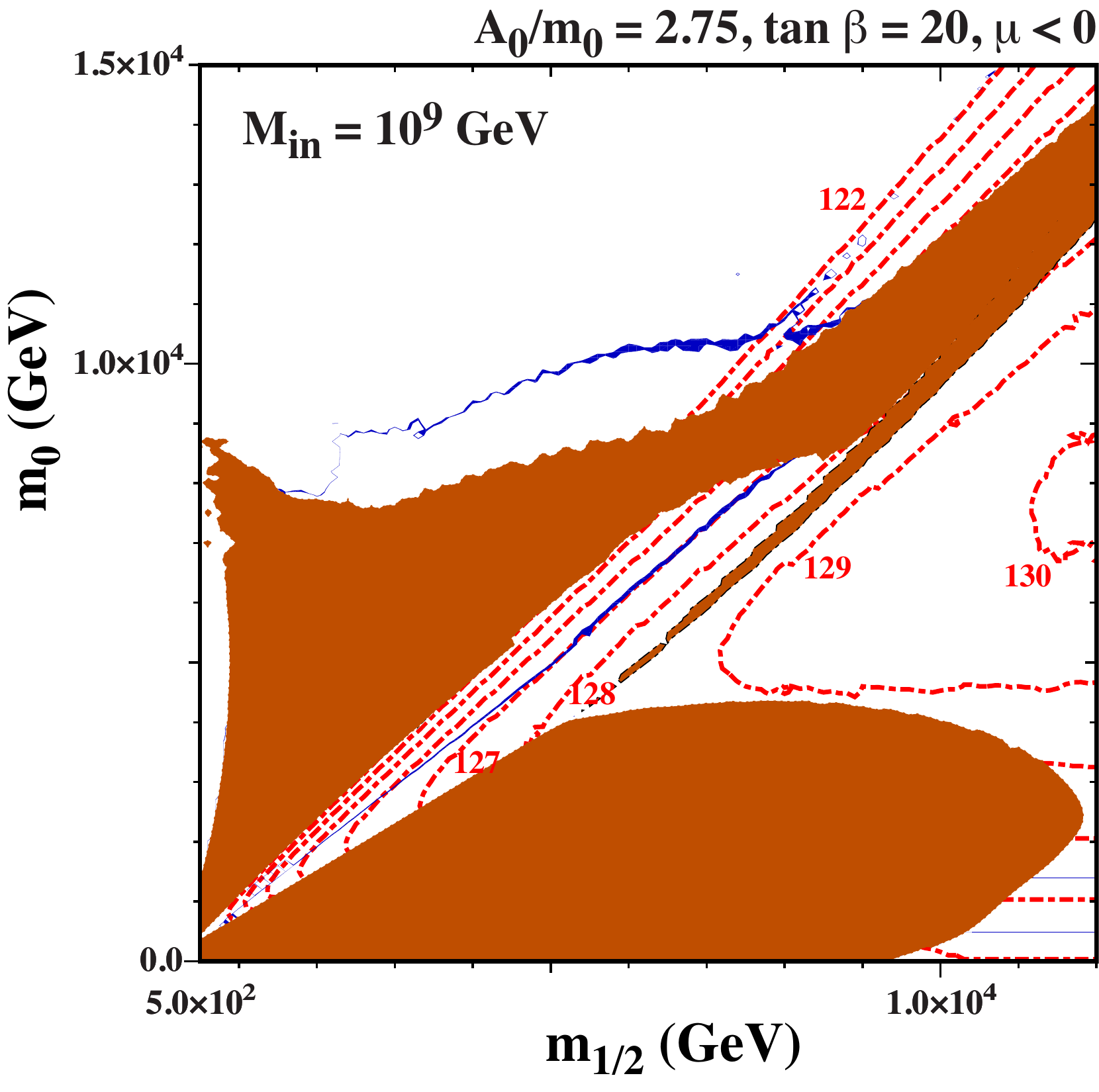}
 \end{minipage}
 \hspace{0.5cm}
  \begin{minipage}[b]{0.4\textwidth}
     \includegraphics[width=\textwidth]{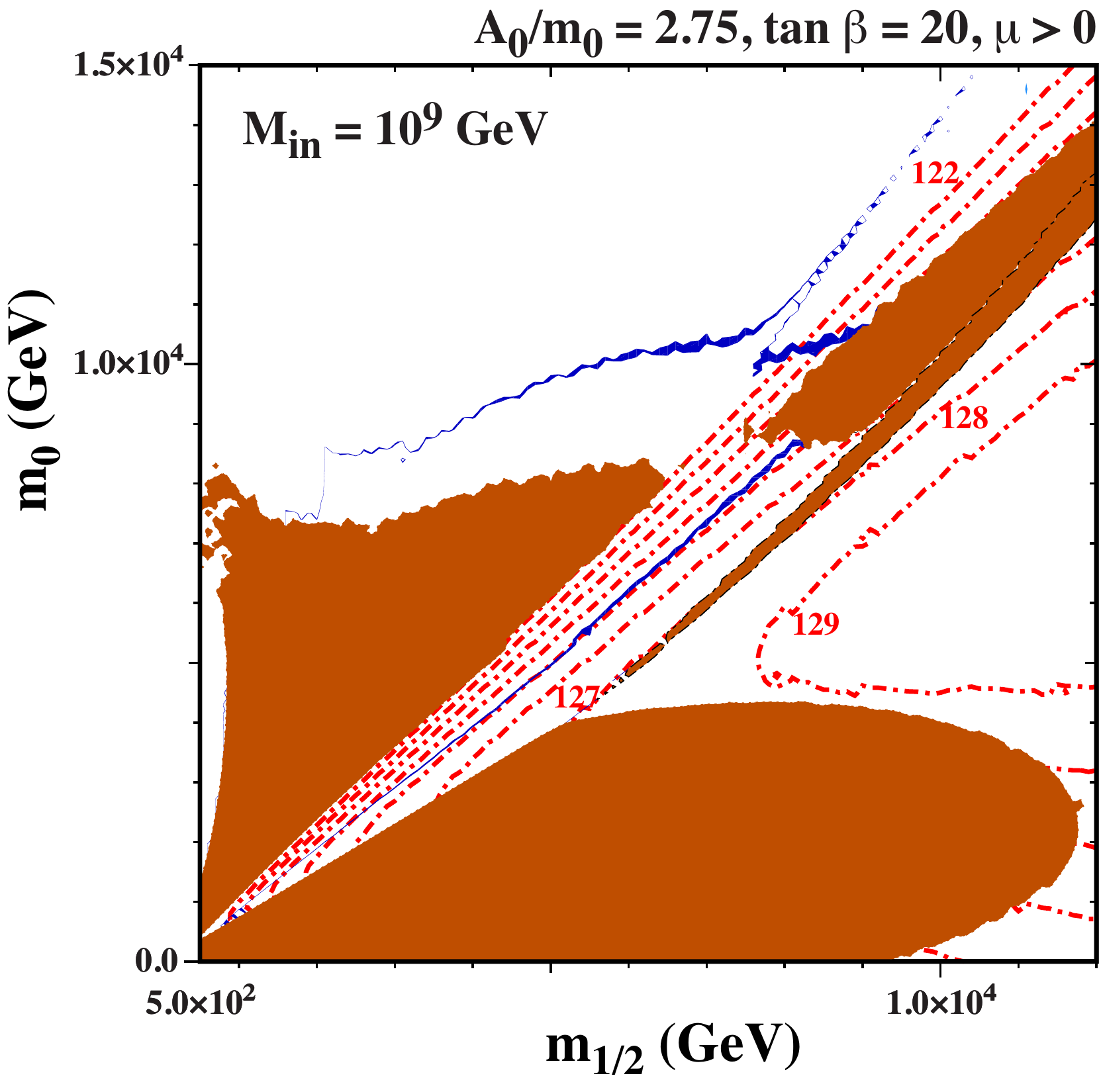}
 \end{minipage}
\caption{\it The $(m_{1/2},m_0)$ planes in a sub-GUT model with $M_{in}=10^9$ GeV and $\tan \beta = 20$.
In the left panel,  $A_0 = 2.75 m_0$ and $\mu < 0$, whereas in the right panel
$A_0=2.75 m_0$ and $\mu>0$. Strips with the allowed cosmological LSP density are
shaded dark blue. In this figure the 3$\sigma$ Planck range $0.1151 < \Omega h^2 < 0.1235$~\protect\cite{Planck} is used. Regions where the LSP is charged are shaded brick red
and contours of $M_h$ are indicated by red dot-dashed lines.}
\label{fig:sub-GUTplane}
\end{figure}

In Fig.~\ref{fig:sub-GUTplane} we compare two $(m_{1/2},m_0)$ planes in a
sub-GUT model with $M_{in}=10^9$ GeV and  $\tan\beta=20$.
In the left panel,  $A_0 = 2.75 m_0$ and $\mu < 0$ while in the right panel
$A_0=2.75 m_0$ and $\mu>0$.
% The stop-coannihillation strip close to this region is quite narrow,
%so we have enhanced its visibility by shading a larger range of relic density than that allowed by cosmology: $\Omega_\chi h^2 \in (?, ?)$,
%while indicating by a cross the tip of the strip for the allowed density range.
As in Fig. \ref{fig:cmssmplanes}, the brick-shaded regions are where the LSP is charged.
Here,
the lighter stop squark is the LSP in the brick-coloured region at small $m_{1/2}$ and relatively large $m_0$,
as well as in a diagonal band extending to large $m_{1/2}$ and $m_0$.
%from $(m_{1/2}, m_0) \sim (8, 9)$~TeV to (12, 13)~TeV and beyond.
In the right panel, the stop LSP region is split in two parts with the area between them
having a mixed Higgsino/Bino LSP.
There is a narrower band extending from $(m_{1/2}, m_0) \sim (6, 5)$~TeV to (12, 13)~TeV and beyond, outlined in black,
where the LSP is a charged Higgsino. The chargino LSP arises because the LSP is changing from being Bino-like
in the upper left corner of the figure to Higgsino-like in the lower right corner. When the Bino and Higgsino masses become degenerate,
the mixing becomes significant, lifting the masses of the neutral Higgsinos and leading to a charged Higgsino LSP.
Finally, at large $m_{1/2}$ and relatively small $m_0$ there is a region where the LSP is the lighter stau.

Above and to the right of the stop LSP region in the right panel of Fig.~\ref{fig:sub-GUTplane} there is a pair of blue
bands where the relic neutralino LSP density falls within the range indicated by the Planck and other measurements \cite{Planck}.
These aim towards a vertex at $m_{1/2} \sim m_0 \sim 10$~TeV that lies, however, within the Higgsino LSP region
that separates the two bands. We note in addition  the appearance of an outwards-pointing spike in the upper dark matter strip, whose base is at
$(m_{1/2}, m_0) \sim (8, 10)$~TeV. Along the two flanks of this spike, the dark matter density is
brought down into the range allowed by cosmology by rapid ${\tilde t_1} {\tilde t_1^*}$
annihilation via the heavy Higgs bosons $H$ in the direct channel~\footnote{The actual extent of the spike is unclear due to numerical
uncertainties in calculating the Higgs bosons mixing angle.}. Compared with the stop coannihilation strips in the CMSSM, these subGUT
strips are relatively thick and clearly visible in the figure, even with the more restrictive range shown 
for the dark matter relic density. Unlike the CMSSM, the mass parameters along the stop 
coannihilation strip in sub-GUT models are much less tuned.  In the upper part of the strip (which is nearly horizontal with $m_0 \sim 10$ TeV),
the difference between the stop and neutralino mass changes very slowly with increasing $m_0$.
As a result there is a broad region between the strip and the stop LSP shaded region where the relic density is too small
to account for all of the dark matter. Because the difference between the stop and neutralino mass varies slowly
in this figure, the coannihilation strip is relatively thick, and here we have plotted the 3 $\sigma$ Planck range rather than
the extended ranges used in the other figures.

Also shown in Fig.~\ref{fig:sub-GUTplane} as dot-dashed red lines are contours of $M_h$ calculated using {\tt FeynHiggs 2.13.0}.
Taking into account the uncertainty in this calculation, we see that the narrower, lower, diagonal part of
the stop coannihilation strip that extends from low $(m_{1/2}, m_0)$ towards the charged Higgsino LSP region
is all compatible with the LHC measurement of $M_h$. Some of the stop coannihilation
strip between the Higgsino LSP region and the outwards-pointing spike may also be
consistent with the Higgs mass, given the uncertainties in the calculations.
However, for $\mu >0$, along the spike and in the region to the left of the spike, $M_h$ appears to be too small.
The situation is improved for $\mu < 0$ as the spike may be compatible, but much of the horizontal strip still lies
at low $M_h$.
Nevertheless, since there is a great deal of uncertainty in the Higgs mass calculation,
as discussed in the previous Section, we cannot exclude the possibility that
some portions of these parts of the stop strip might also yield an acceptable Higgs boson mass.

\begin{figure}[ht!]
  \centering
\includegraphics[width=.65\textwidth]{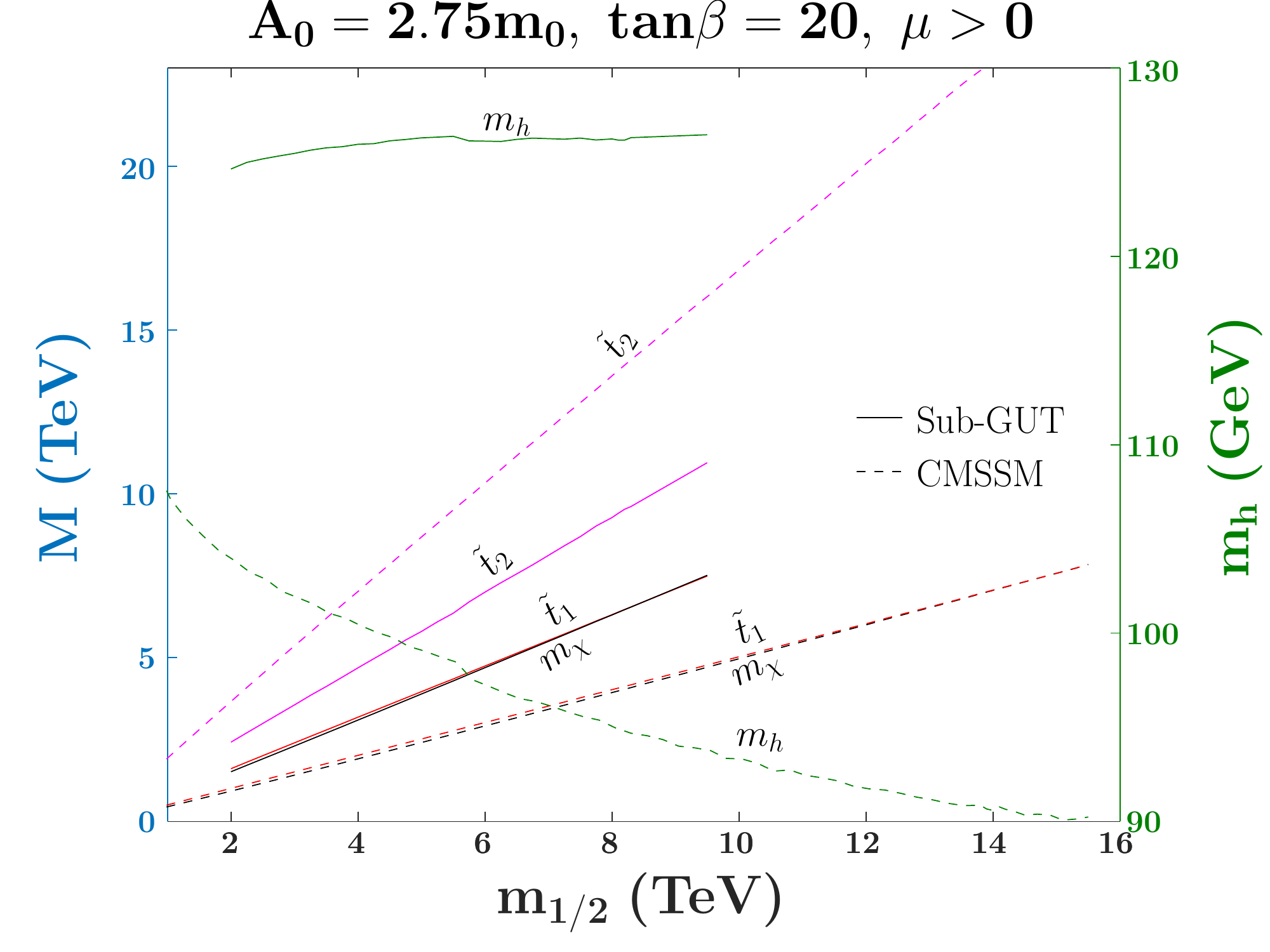}
\caption{\it The masses of the neutralino LSP, the left and right stop, and the SM like Higgs boson in the CMSSM  and sub-GUT with $M_{in}=10^9$ GeV for $A_0=2.75 m_0$, $\tan\beta=20$ and $sgn(\mu)>0$. }
\label{fig:CompCMSSMSubGut}
\end{figure}

We show in Fig.~\ref{fig:CompCMSSMSubGut} a comparison of relevant masses,
plotted as functions of $m_{1/2}$, in the CMSSM and a sub-GUT model
with $M_{in}=10^9$ GeV, both with $A_0=2.75 m_0$, $\tan\beta=20$, and $\mu>0$.
The LSP mass lines are black, those for the $\tilde t_1$ are red, and those for the $\tilde t_2$ are purple.
The dashed lines are for the masses in the CMSSM  and the solid lines are for the sub-GUT model. In the sub-GUT case, we only plot the portion of the strip which is in between the stop and stau LSP regions.  Since $m_\chi$ and $m_{\tilde t_1}$ depend mostly on $m_{1/2}$, their masses along the other portion of the strip just overlap the existing lines. The Higgs masses on the missing portions of the strip can be inferred from  Fig.~\ref{fig:sub-GUTplane} and tends to be too small while $m_{\tilde t_2}$ is slightly larger.
In both the CMSSM and sub-GUT planes, the LSP, $\chi$, and $\tilde t_1$ are nearly degenerate, whereas $\tilde t_2$ is
significantly heavier. The lines are truncated at the tips of the stop strips, and the masses can be read off from the left vertical axis.
As can be seen in the Figure, the Bino LSP of the sub-GUT model is much heavier than its CMSSM counterpart for the same $m_{1/2}$,
but the maximum LSP masses are similar in the two cases because the strip is longer in the CMSSM case,
both being $\sim 7$~TeV. We also see that the
stop masses of the sub-GUT are much less split than in the CMSSM, which leads to a greatly enhanced Higgs boson mass~\footnote{In the CMSSM for $A_0<0$, the Higgs mass is considerably better along the stop coannihilation strip but the extent is drastically reduced due to the smaller
value of $|A|$ at the SUSY scale.}.
For this reason, the Higgs boson mass constraint completely rules out the coannihilation strip for the CMSSM
(see the dashed green line),
but places no meaningful constraints on the sub-GUT model. The solid green line shows that $M_h$ is rather insensitive to
$m_{1/2}$ along the sub-GUT coannihilation strip, with a mass (to be read off the right vertical axis) that is
compatible with 125~GeV within the calculational uncertainties.

\begin{figure}[th!]
\centering
 \begin{minipage}[b]{0.47\textwidth}
    \includegraphics[width=\textwidth, ]{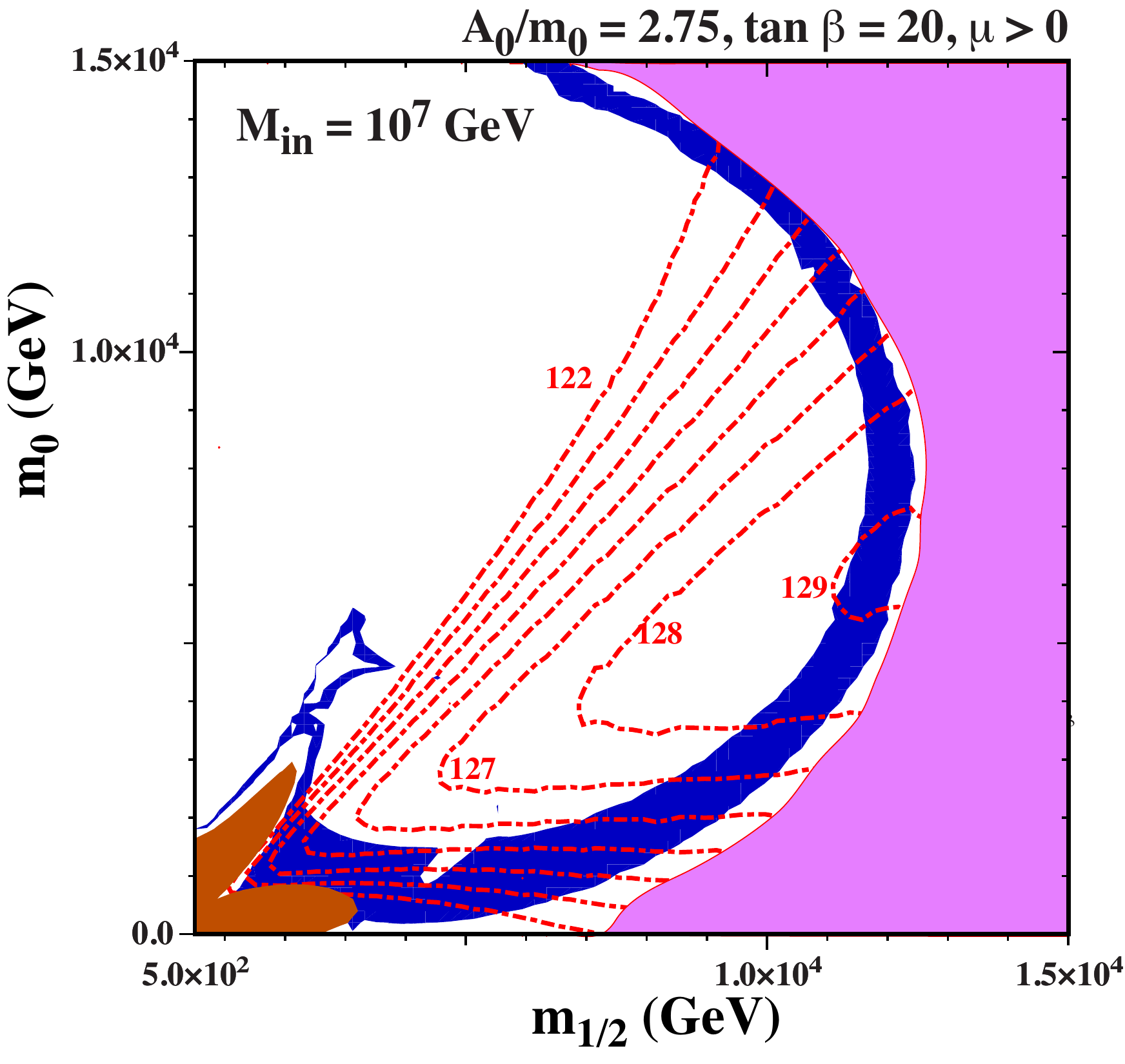}
 \end{minipage}
\hspace{0.5cm}  \begin{minipage}[b]{0.47\textwidth}
     \includegraphics[width=\textwidth ]{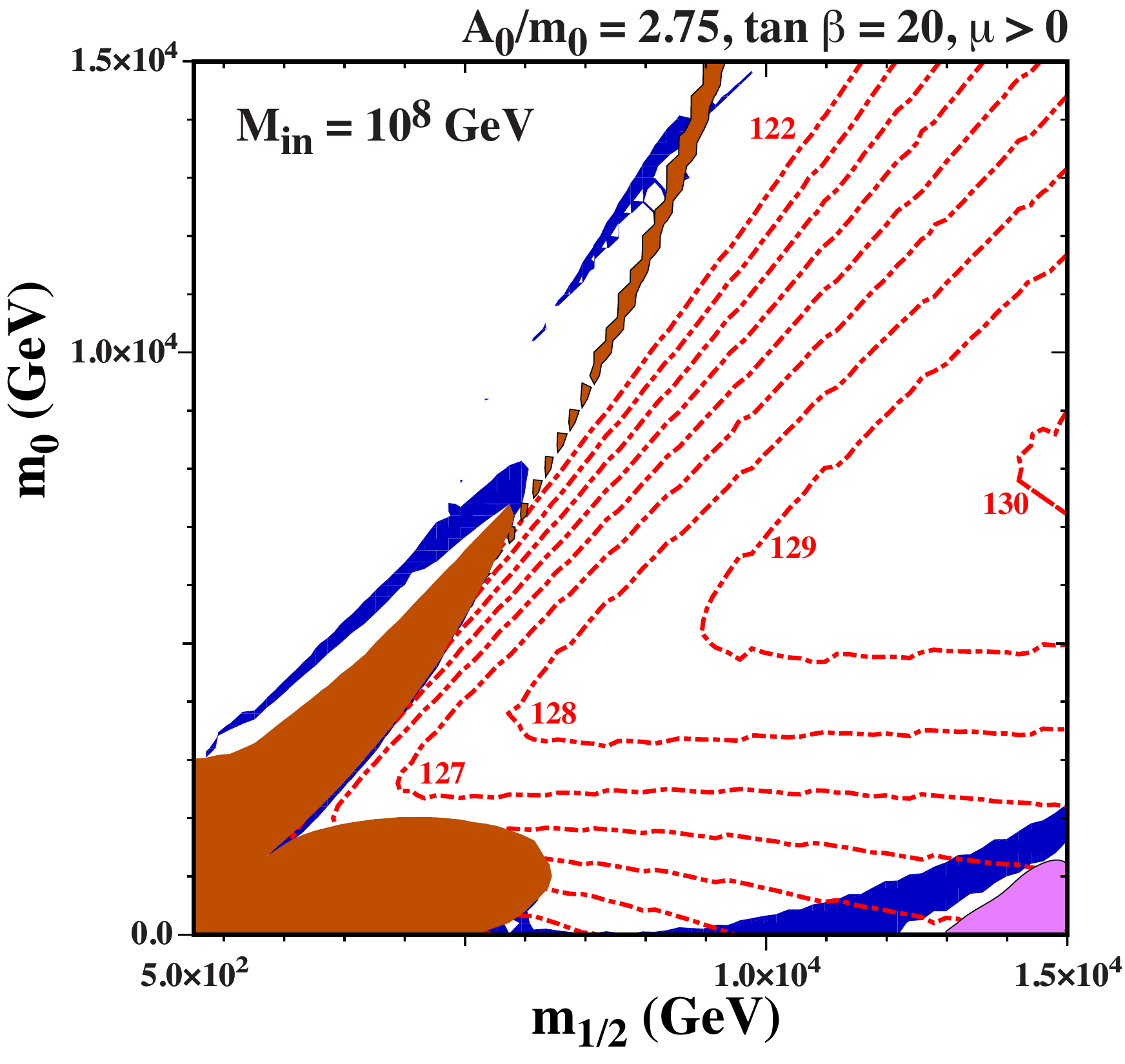}
 \end{minipage}
   \begin{minipage}[b]{.47\textwidth}
     \includegraphics[width=\textwidth ]{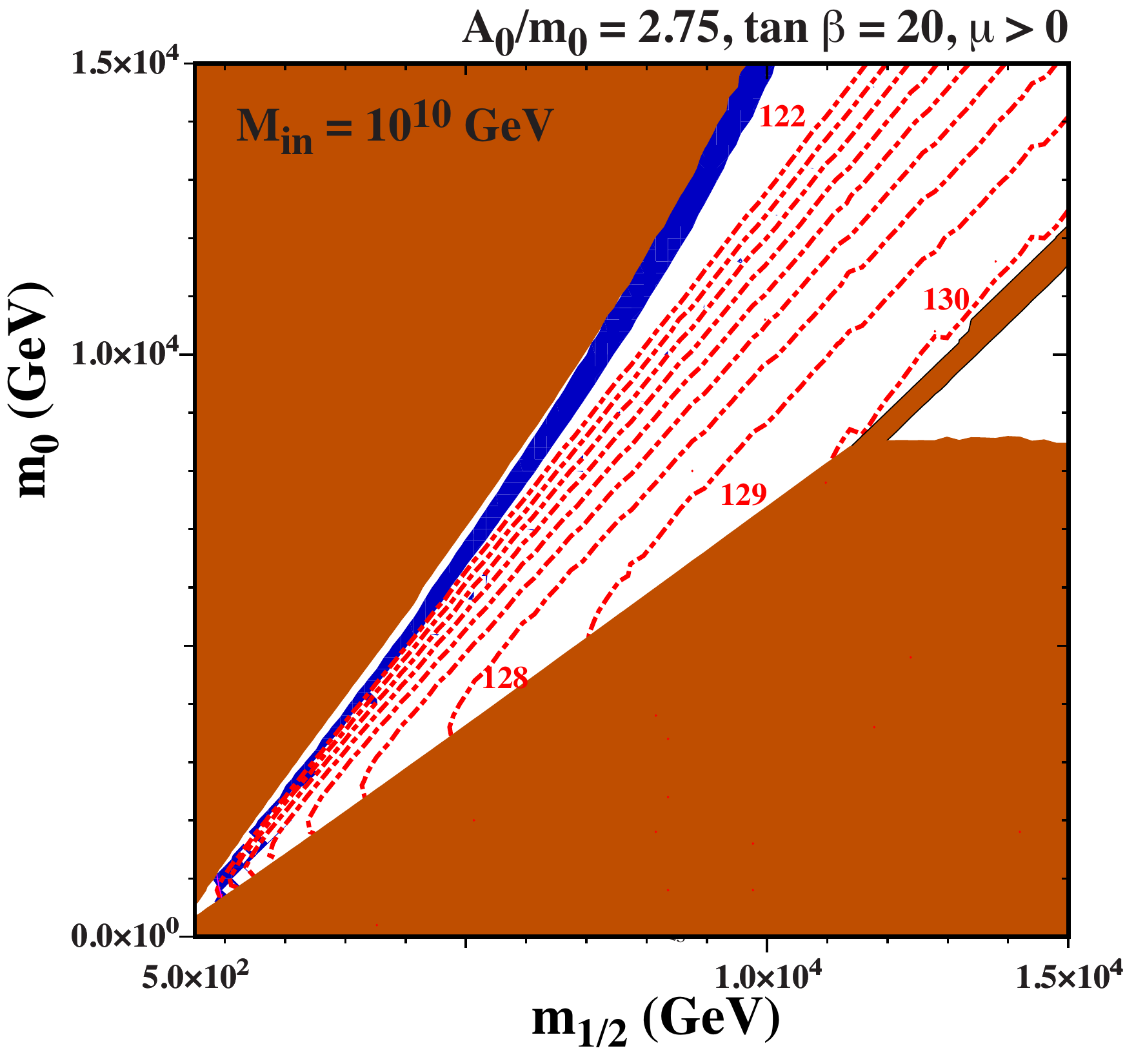}
 \end{minipage}
\hspace{0.5cm}
  \begin{minipage}[b]{.47\textwidth}
     \includegraphics[width=\textwidth ]{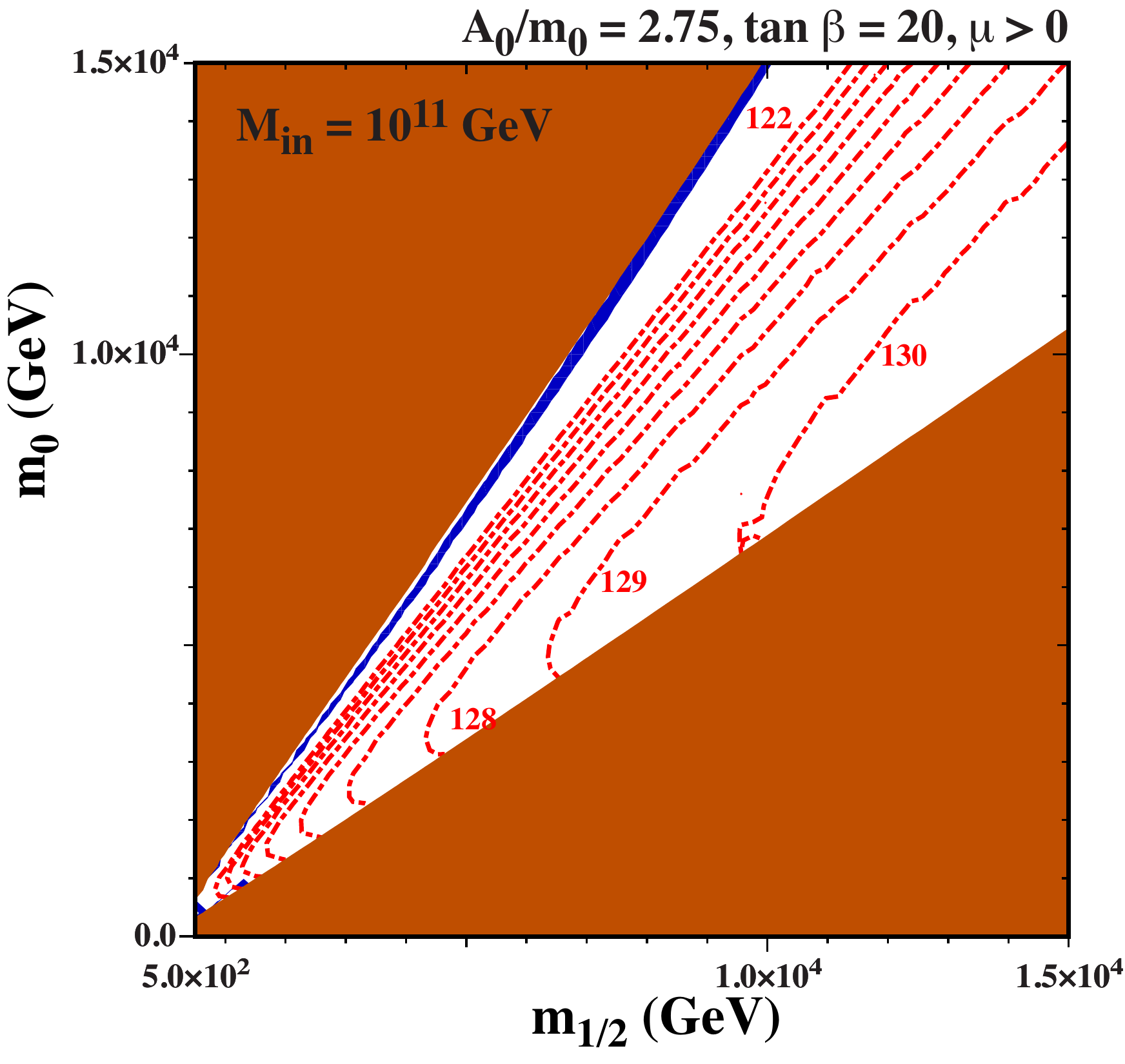}
 \end{minipage}
\caption{\it The $(m_{1/2},m_0)$ planes sub-GUT models with
$\tan \beta = 20$, $A_0 = 2.75 \, m_0$ and $\Min = 10^{7}$~GeV (upper left), $\Min=10^{8}$ GeV (upper right),
$\Min=10^{10}$ GeV (lower left), and $\Min=10^{11}$ GeV (lower right).
The pink region is where $\mu^2<0$ and radiative EWSB is not possible. Other shadings used here are the same as Fig.~\ref{fig:sub-GUTplane}.
The shading for the relic density is enhanced so that $0.06 < \Omega h^2 < 0.2$.
}
\label{fig:sub-GUTplaneMin}
\end{figure}

We examine in Fig.~\ref{fig:sub-GUTplaneMin} the impact of changing $\Min$ on the $(m_{1/2},m_0)$
plane in Fig.~\ref{fig:sub-GUTplane}. We fix $\tan\beta=20$ and $A_0=2.75 \, m_0$ and
choose $\Min=10^x$ GeV where $x=7,8,10,11$ for the upper left, upper right, lower left and lower right panels,
respectively.  In all the panels of Fig.~\ref{fig:sub-GUTplaneMin} the brick-shaded region adjacent to the $m_0$ axis
corresponds to a stop LSP, whilst in the brick-shaded region adjacent to the $m_{1/2}$ axis the LSP is a stau.
For $x=8,10,11$, in the brick-shaded region that is outlined in black the LSP is a charged Higgsino. In the upper
panels, electroweak symmetry breaking (EWSB) is not possible in the pink regions at large $m_{1/2}$.
For $\Min=10^7$ GeV, in the region with large $m_{1/2}$, $\Min$ is so low that the renormalization-group
running is insufficient to drive the up-type Higgs soft mass negative for large gaugino masses, so there electroweak symmetry breaking (EWSB)
does not occur. The dark matter strip around the brick-shaded region that is adjacent to the $m_0$ axis is due to stop coannihilation,
and the portion of the strip near the no-EWSB region has a typical Higgsino thermal relic with $\mu\sim 1.1$ TeV.
For $\Min=10^8$ GeV, the stop LSP region becomes larger, extending the stop coannihilation strip to larger $m_{1/2}$
and $m_0$. In this case the no-EWSB region moves to larger $m_{1/2}$, and the region of
parameter space with a Higgsino LSP shrinks.

Besides the strip-like regions, there is also a blue ring shape region on top of the stop strip in the panel for $M_{in} = 10^7$~GeV and a blue sliver that intersects with the chargino LSP strip seen in the panel for $M_{in} = 10^8$~GeV. In these regions of parameter space,
the masses of the three lightest neutralinos are quite similar. Because of this, the $\chi_1$, which is mostly a Bino, and the $\chi_2$ and
$\chi_3$, which are mostly Higgsinos, can coannihilate, with an enhancement from the heavy Higgs funnel~\footnote{Unlike the
heavy Higgs funnel in Fig.~\ref{fig:sub-GUTplane}, the extent of this funnel is not sensitive to the exact value of the Higgs mixing angle.}.
In the two panels, which have $\Min=10^{7,8}$~GeV, the renormalization-group
running is insufficient to give a stop LSP unless $m_{1/2}$ is relatively small. Because of this, the extent of the stop
coannihilation strip is greatly reduced. The maximum $m_\chi$ of the stop strips are $\sim2$~GeV and $5$~GeV for $\Min=10^{7,8}$~GeV respectively. In the panel with $\Min=10^{10}$ GeV, the stop LSP region has grown significantly,
and is accompanied by a stop coannihilation strip that extends beyond the displayed part of the plane. The stau LSP region
levels off at $m_{1/2}\sim 11$ TeV.  For $\Min=10^{11}$ GeV,  the plane has become qualitatively similar to that in
the CMSSM. For $\Min=10^{10,11}$ GeV, the mass splitting of the stops has become large enough that the
Higgs mass is suppressed to such an extent that the stop strip no longer has a viable Higgs mass and so is
excluded~\footnote{For $\mu<0$, the value of the Higgs mass is improved, but is still much too small to meet experimental constraints.}.
For $\Min=10^{9}$ GeV, as seen already in Fig.~\ref{fig:sub-GUTplane}, we are between these two extremes,
and the stop LSP region is large enough to give a coannihilation strip that extends to large $m_{1/2}$,
but is not so large that the Higgs boson mass becomes too small. In this case, the maximum value of $m_\chi$ in the stop strip after taking $M_h$ into account is $\sim7.4$~TeV, lying on top of the diagonal brick-shaded band in the right panel of Fig~\ref{fig:sub-GUTplane}.

\begin{figure}[ht!]
\centering
 \begin{minipage}[b]{0.47\textwidth}
    \includegraphics[width=\textwidth ]{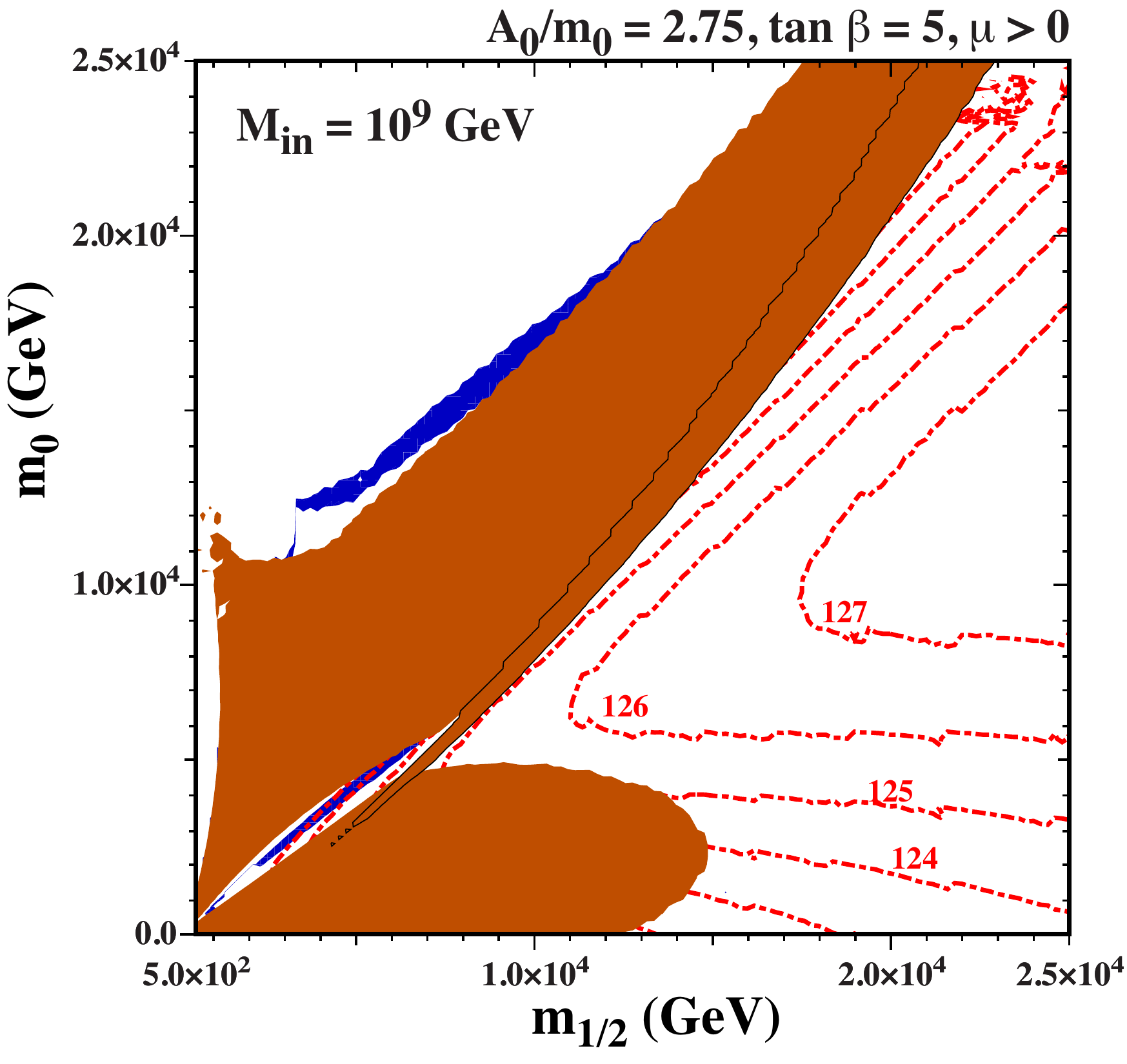}
 \end{minipage}
\hspace{0.5cm}  \begin{minipage}[b]{0.47\textwidth}
     \includegraphics[width=\textwidth ]{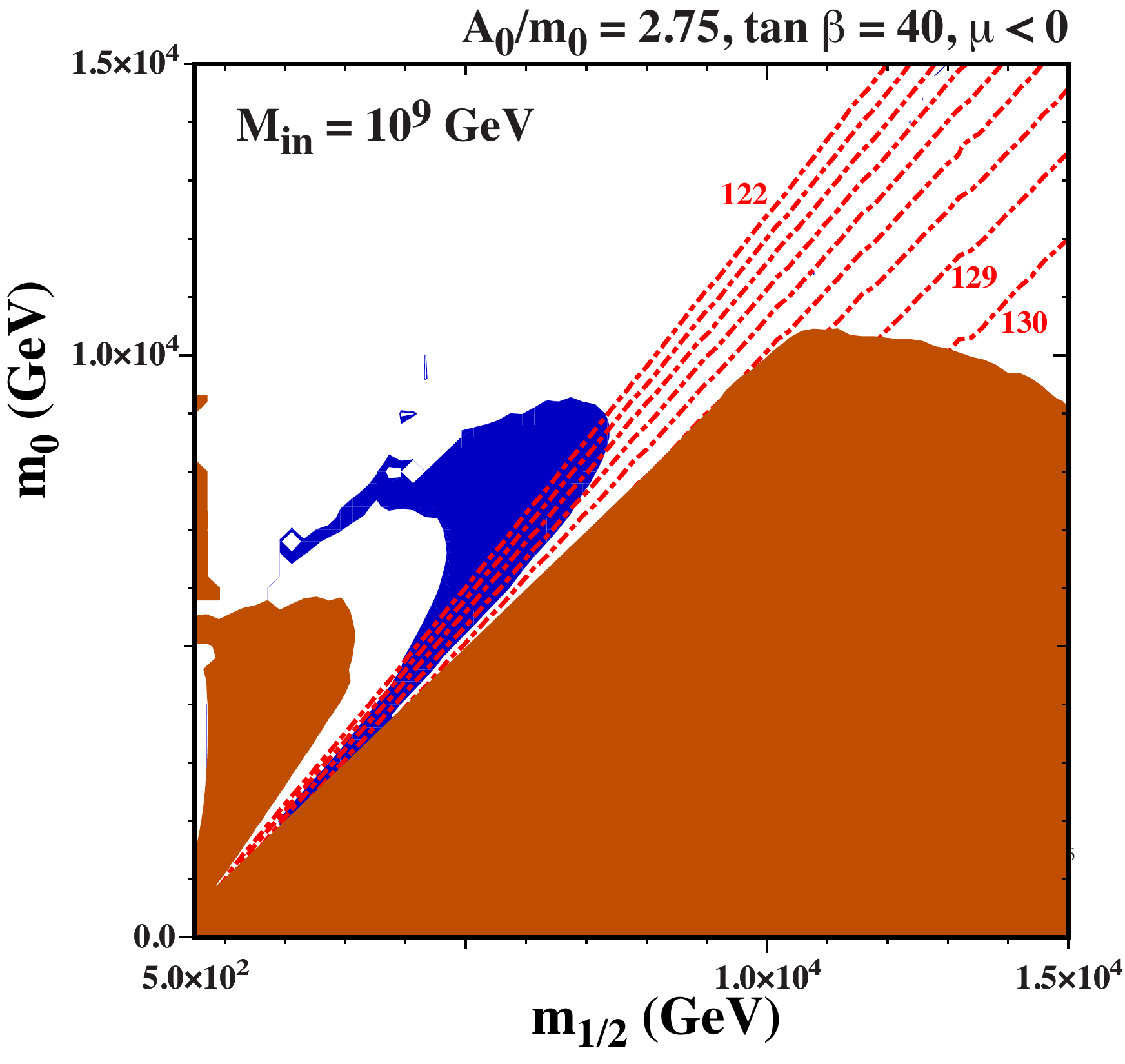}
\end{minipage}
\caption{\it The $(m_{1/2},m_0)$ planes for sub-GUT models with
$\Min=10^9$ GeV and $A_0 = 2.75 \, m_0$, for $\tan\beta = 5$ (left panel) and $\tan\beta=40$ (right panel).
The shadings are the same as in Fig.~\ref{fig:sub-GUTplaneMin}.}
\label{fig:sub-GUTplanetanb}
\end{figure}

The next pair of plots examines the effect of changing $\tan\beta$. In Fig.~\ref{fig:sub-GUTplanetanb}
we show the $(m_{1/2},m_0)$ planes for $\Min=10^9$ GeV and $A_0=2.75 \, m_0$ for  $\tan\beta=5 (40)$
in the left (right) panel. In the case with $\tan\beta=5$ the stop LSP region has grown because of the larger
top Yukawa coupling. Indeed, it has grow so much that it merges with the charged Higgsino LSP region,
forcing the lower stop coannihilation strip to terminate at a much lower value of $m_{1/2} \sim 7$~TeV and $m_\chi\sim 5.4$~TeV.
Within this region, $M_h$ is compatible with 125~GeV, within the calculational uncertainties.
The upper stop coannihilation strip extends farther in $m_{1/2}$ and $m_0$, but the Higgs mass is very low here~\footnote{The
Higgs mass is again improved for $\mu<0$, but is still far below the experimental constraint.}.
For $\tan\beta=40$,
on the other hand, the stop LSP region has shrunk while the stau LSP region has grown. This is again due the
$\tan\beta$ dependence of the Yukawa couplings. The increase in the tau Yukawa coupling is responsible
for the larger stau LSP region, and the smaller top Yukawa coupling for the smaller stop LSP region. The
shrinking of the stop LSP regions leads to the stop coannihilation strip terminating at much lower values of
$m_{1/2}\sim 7$~TeV and the LSP mass $\sim 5.5$~TeV~\footnote{Although a large portion of the coannihilation strip follows the stau LSP region,
the relic density is reduced through coannihilation with the stop, which also has a similar mass to the LSP.},
compared with the case of a more moderate value of $\tan\beta = 20$
studied previously, which leads to a stop coannihilation strip that extends to larger LSP masses.

\begin{figure}[ht!]
\centering
 \begin{minipage}[b]{0.47\textwidth}
    \includegraphics[width=\textwidth ]{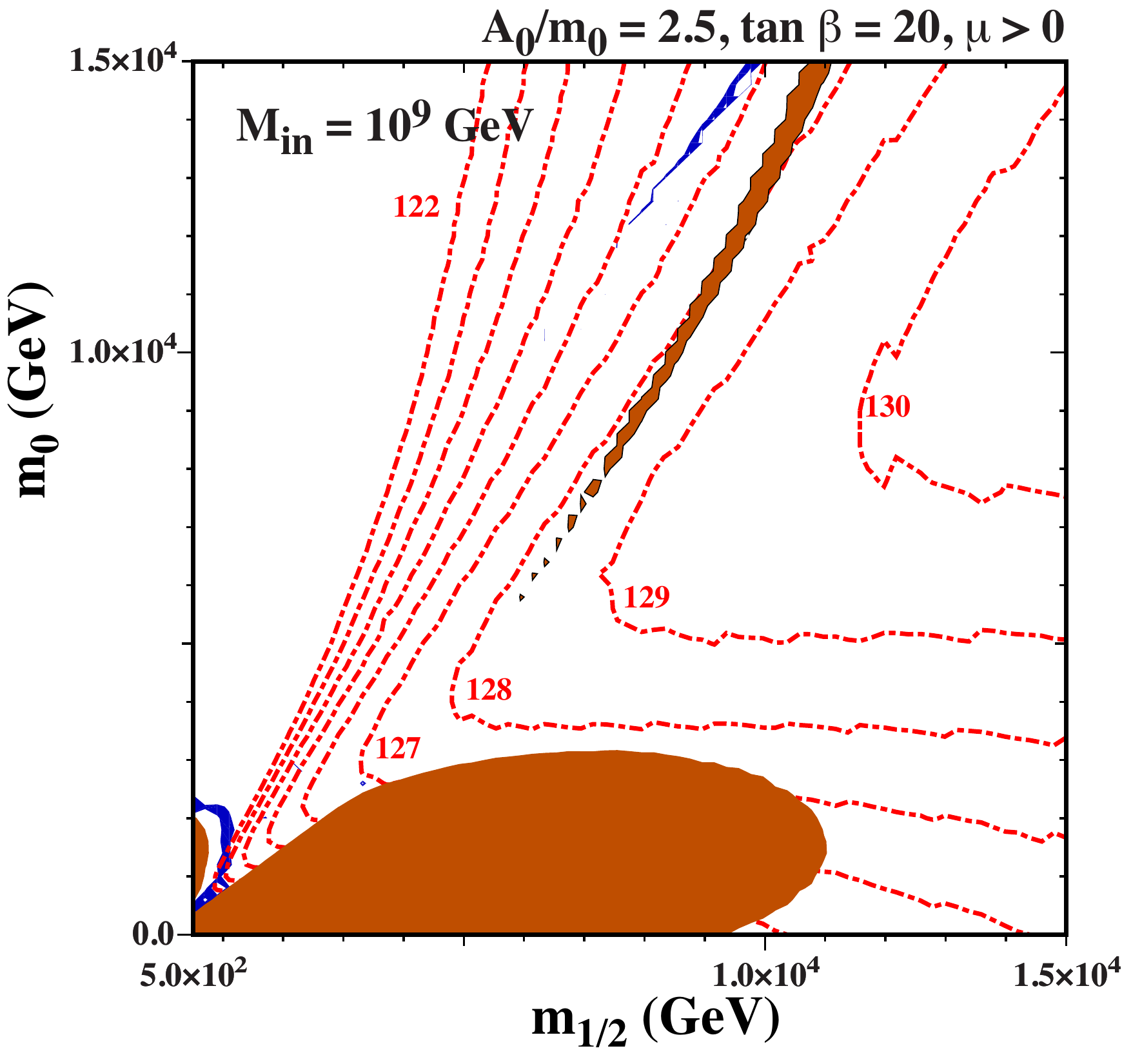}
 \end{minipage}
\hspace{0.5cm}  \begin{minipage}[b]{0.47\textwidth}
     \includegraphics[width=\textwidth]{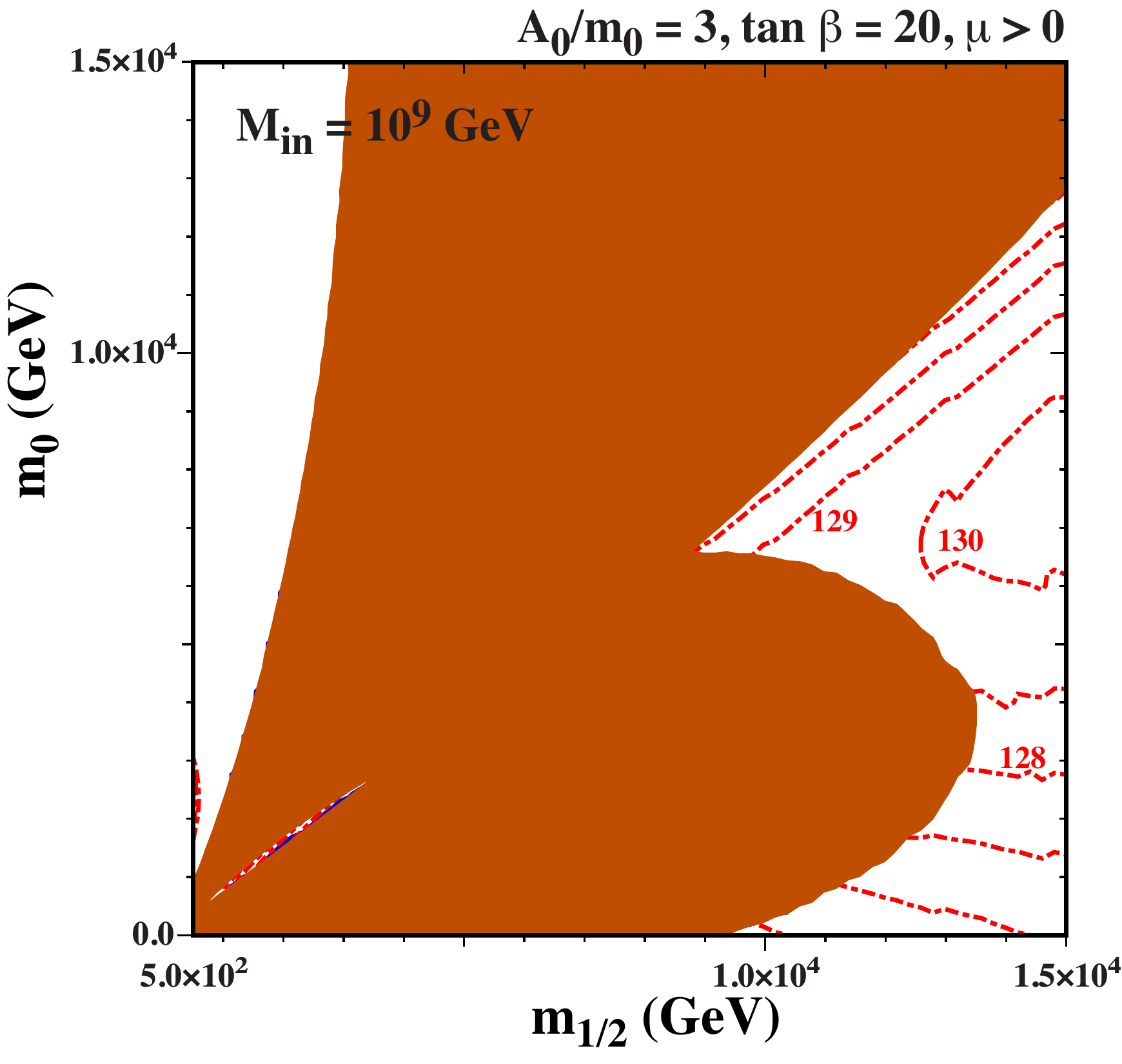}
\end{minipage}
\caption{\it The $(m_{1/2},m_0)$ planes for sub-GUT models with
$\Min=10^9$ GeV, $\tan\beta=20$ and for $A_0 = 2.5 \, m_0$ (left panel) and $A_0=3 \, m_0$ (right panel).
The shadings are the same as in Fig.~\ref{fig:sub-GUTplaneMin}.}
\label{fig:sub-GUTplaneA0}
\end{figure}

The last pair of plots show the effect of varying $A_0$.  In Fig.~\ref{fig:sub-GUTplaneA0} we show the
$(m_{1/2},m_0)$ planes for $\Min=10^9$ GeV and $\tan\beta=20$ for $A_0=2.5 \, m_0$ ($A_0=3 \, m_0$) in the left (right) panel.
In the left panel, the smaller $A$-term is unable to push the stop mass tachyonic unless $m_{1/2}$ is very small.
Because of this, the stop coannihilation strip has all but disappeared,
clinging on only when $m_{1/2} \lesssim 2$~TeV in a region where the value of $M_h$ calculated using
{\tt FeynHiggs~2.13.0} is $< 125$~GeV. As was the case in Fig.~\ref{fig:sub-GUTplaneMin},
the blue sliver in the upper part of the panel appears because $\chi_1$, $\chi_2$, and $\chi_3$ are nearly degenerate
and have an enhanced annihilation rate to the (nearly) on-shell heavy Higgs bosons. For a larger $A$-term, as shown in
the right panel, the stop coannihilation region grows and merges with the stau coannihilation region and
charged-Higgsino strip. The parameter space no longer exhibits a stop coannihilation strip.
As one can see, the spectrum at $M_{in} = 10^9$ GeV, is very sensitive to $A_0/m_0$,
and the rich structure seen in the previous figures requires $A_0/m_0 \approx 2.75$.

\section{Conclusions}

It is well-known that an effective way to reduce the relic density of a massive LSP into the range
allowed by Planck and other astrophysical and cosmological observations is coannihilation of the LSP
with at least one other particle of similar mass that decouples at around the same time in the early universe. In this case, as long as the temperature of the thermal
bath is no smaller than the difference between the particle masses, the lighter one - the LSP
candidate - can scatter in the thermal bath and be converted to the heavier particle.
If the heavier particle can annihilate efficiently into Standard Model particles, the relic density
of the dark matter candidate can be significantly reduced. The more strongly interacting the coannihilating particle,
the more efficient it is at reducing the relic density.  In supersymmetric models, one of the most effective
coannihilating partners for the LSP is the lighter stop. It is particularly effective because stop-antistop
annihilation rates to $hh$ and $WW/ZZ$ are enhanced if the $A$-terms are large.
This enhancement of the annihilations to $WW/ZZ$ arises because the longitudinal
Goldstone bosons interact with the stops through $A$-terms. Specifically, the amplitudes
for $\tilde t \tilde t^*\to WW,ZZ,hh$ all receive an enhancement proportional to
$A_t^2/(m_{\tilde t_R}^2+m_{\tilde t_L}^2)$. On top of this enhancement,
these annihilation processes are boosted by Sommerfeld enhancement and by bound state formation.

In the CMSSM, these enhanced annihilation rates of stops allow the relic density
to be consistent with Planck constraints for $m_{1/2} \lesssim 17$ TeV, with an LSP mass
$m_{\chi}\sim 8.5$ TeV. However, a Bino mass this large requires $A_0 \sim 5 \, m_0$.
Such a large $A$-term splits the stop mass eigenstates severely,
leading to an unacceptably small mass for the lightest supersymmetric Higgs boson.
In the CMSSM, the Higgs mass measurement places a strong constraint on the
extent of the stop coannihilation strip, if the Higgs mass is calculated with {\tt FeynHiggs 2.13.0},
the supersymmetric Higgs mass calculator we use for this study. With the Higgs mass constraint included,
$A_0 \sim 5 \, m_0$ is no longer viable. For $A_0/m_0=-4.2$, $\tan\beta =5$ and $\mu>0$, the stop coannihilation strip can only reach
$m_{1/2} \sim 7.2{~\rm TeV}$ with an LSP mass of $m_{\chi}\sim 3.5{~\rm TeV}$ for $m_0\sim 12{~\rm TeV}$,  

However, it is not clear how reliable Higgs mass calculators are in this extreme regime,
and therefore how seriously one should take this constraint on the stop coannihilation strip.
This concern arises from the fact that the various publicly available Higgs mass calculators
yield very different results in this regime. For the CMSSM with $A_0=5 \, m_0$, $\tan\beta=20$
and $\mu>0$, the Higgs mass calculators give a mass for the Standard Model-like Higgs boson
that span a range of order 70 GeV. Although the spread is smaller for $A_0<0$,
where the Higgs mass spans a range of order 10 GeV for $A_0=-4.2m_0$,
$\tan\beta=20$ and $\mu>0$, this regime is much less extreme and the extent of the stop
coannihilation strip is drastically reduced. This is due to the fact that for $A_0<0$,
the renormalization-group running suppresses the $A$-terms much more drastically
as they are run down to the supersymmetry-breaking scale. With such smaller $A$-terms,
the Higgs mass calculations become much more reliable and the annihilations
$\tilde t \tilde t^* \to hh,WW,ZZ$ becomes much less effective. Because of this,
the stop coannihilation strip only reaches $m_{1/2}\sim 7$ TeV and $m_{\chi} \sim 3.5$~TeV
for $A_0=-4.2 \, m_0$, $\tan\beta=20$ and $\mu>0$.
Thus, if the results of {\tt FeynHiggs 2.13.0} in this regime are in fact reliable,
the extent of the stop coannihilation strip in the CMSSM is drastically reduced.

However, for stop masses that are more degenerate, the Higgs mass constraints
are less restrictive.  In sub-GUT models, in which the soft supersymmetry-breaking
masses unify at some lower scale $\Min$, the low-scale stop masses tend to be
more degenerate due to the reduced running. Because of this increased degeneracy,
a Bino LSP mass of $m_\chi \sim 7$ TeV can be consistent with the Planck relic density
measurement for $A_0=2.75 \, m_0$, $\tan\beta=20$, and $\mu>0$ with $\Min=10^9$ GeV.
Moreover, in this case calculations with {\tt FeynHiggs 2.13.0} yield a mass of the
lightest supersymmetric Higgs boson that is compatible with 125~GeV all the way
to the tip of the stop coannihilation strip.

In addition to demonstrating that an LSP mass $m_\chi \sim 8$ TeV can be
compatible with the relevant dark matter density and Higgs mass constraints
in a sub-GUT model, our work highlights the importance of annihilations
into the longitudinal modes of massive gauge bosons, as well as the
Sommerfeld enhancement and bound-state effects. It also highlights the
need for a reliable code to calculate the lightest Higgs mass in the extreme
regions of parameter space that are relevant for large LSP masses. The
results obtained with {\tt FeynHiggs 2.13.0} may well be reliable in many
cases, but corroboration is essential.

\section*{Acknowledgements}

We would like to thank S. Heinemeyer for his continuing help with our implementation of {\tt FeynHiggs}.
The work of J.E. was supported in part by STFC (UK) via the research grant ST/L000326/1,
and in part by the Estonian Research Council via a Mobilitas Pluss grant.
F.L. was supported by the World Premier International Research Center Initiative (WPI), MEXT, Japan.
The work of K.A.O. was supported in part by DOE grant DE-SC0011842 at the University of Minnesota.
The work of J. Z. was supported in part by KAKENHI Grant Number JP26104009.

\end{document}